\documentclass[reprint,amsmath,amssymb,pra]{revtex4-1}
\usepackage{amsmath}
\usepackage{amssymb}
\usepackage{mathrsfs}
\usepackage[colorlinks]{hyperref}
\usepackage{graphicx}
\usepackage{dcolumn}
\usepackage{bm}
\makeatletter

\newcommand{\Rmnum}[1]{\expandafter\@slowromancap\romannumeral #1@}
\makeatother

\begin{document}
\title{Quantifying coherence in terms of the pure-state coherence}
\author{Deng-hui Yu}
\author{Li-qiang Zhang}
\author{Chang-shui Yu}
 \email{ycs@dlut.edu.cn}
 \affiliation{School of Physics, Dalian University of Technology, Dalian 116024, China}%
\date{\today }
\begin{abstract}
Quantifying quantum coherence is a key task in the resource theory of coherence. Here we establish a good coherence monotone in terms of a state conversion process, which automatically endows the coherence monotone with an operational meaning. We show that any state can be produced from some input pure states via the corresponding incoherent  channels.  It is especially found that the coherence of a given state can be well characterized by the least coherence of the input pure states, so  a coherence monotone is established by only effectively quantifying the input pure states. In particular, we show that our proposed coherence monotone is the supremum of all the coherence monotones that give the same coherence for any given pure state. Considering the convexity, we prove that our proposed  coherence measure is a subset of the coherence measure based on the convex roof construction. As an application, we give a concrete expression of our coherence measure by employing the geometric coherence of a pure state. We also give a thorough analysis on the states of qubit and finally obtain series of analytic coherence measures.
\end{abstract}
\maketitle
\section{introduction}
Coherence, as the most fundamental  nature of quantum mechanics, is necessary for almost all the other quantum features, such as entanglement  \cite{entanglement,entangle,e1,e2,e3,op2,e4}, quantum correlation  \cite{discord,cor,correlation,Sun_2017}, nonlocality  \cite{non1,non2,non3,non4,non5}, asymmetry \cite{as,as1,as2} and so on. It also plays an important role in many fields including quantum thermodynamics  \cite{ther1,ther2,ther3,ther4,ther5,ther6}, quantum biology  \cite{qb1,qb,qb2,qb3}, quantum metrology  \cite{qm,m1,m2,m3,m4}, quantum phase transitions  \cite{trans1,trans2,trans3,trans4,trans5} etc. Recently, the resource theory of coherence   \cite{qc2,qrt1} has been well developed based on different free operations \cite{qc1,roof1,free,speak,sio,gc,op1,pio}. It not only provides a strict mathematical framework to effectively quantify coherence \cite{qc1,alter,qc2}, but also establishes a  platform to understand quantum mechanical feature in a different perspective.

Up to now, a lot of methods have been proposed to quantify quantum coherence. The most intuitive method could be the coherence measure based on the distance \cite{qc1,trace,multi,ft,cpt} between the state of interest and the closest incoherent state since the corresponding incoherent operations and incoherent states can be unambiguously defined. The remarkable examples are the coherence measure based the $l_1$ norm and quantum relative entropy \cite{qc1}. However, it has been shown that the strong monotonicity in the resource theory requirements has ruled out many convenient norms like the trace norm and other $l_p$ norm ($p\neq 1$) \cite{lp}. In addition, the usual applications of the commutation like the skew information and the Tsallis relative $\alpha$ entropy serve as good coherence measures \cite{i1,i2,i3,i4,i5,i6}. The distinguished feature of the above coherence measures is that they can be analytically calculated for a general state. Besides, the relative entropy coherence has the obvious operational meaning due to its connection with the optimal rate for distilling a maximally coherent state from given states \cite{op1}, the coherence based on the skew information can be related to the quantum metrology \cite{i1,i5}, and the robustness of coherence is shown to be able to describe the advantage enabled by a quantum state in a phase discrimination task\cite{op3}.
The convex roof construction,  a traditional and effective method in the quantification of entanglement measure \cite{eroof,eroof1,monotone}, can also be used to quantify coherence \cite{roof4,roof1,roof2,roof3,op1,oto1}. It is obvious that different quantifications not only provide the different computability, but also imply different operational meanings. How to explore the new understanding of coherence has been still a significant and attractive topic in the resource theory \cite{qrt3,mre,oneshot}.

In this paper, we present the coherence monotone and coherence measure from a completely new perspective. We consider that some pure states undergo incoherent channels \cite{qi2} and finally become the common objective state. It is shown that the coherence of the objective state can be well described by the least coherence of the input pure states. Given any certain coherence monotone $F$ defined on pure states, the coherence monotone extended to mixed states through our method serves as the supremum of all the coherence monotones equal to $F$ for pure states. Considering the convexity, we prove that our coherence measure is a particular subset of the coherence measure based on the convex roof construction. As an application, we select the geometric coherence measure \cite{op2} as the measure for pure states and finally establish the coherence measure for a general states. In particular, we thoroughly analyze the states of qubit. We find the optimal pure state, give the easier method to choosing the coherence measure of pure states, and finally find out the analytic coherence measure for a general quantum state of qubit. This paper is organized as follows. In Sec. II, we elucidate how to describe the coherence based on our incoherent operation process and establish the corresponding coherence monotone. In Sec. III,  we consider the convexity of the coherence monotone and show the connection with the coherence measure based on the convex roof construction. In Sec. IV, we consider the geometric coherence measure as the pure-state coherence to establish our coherence measure. In Sec. V, we thoroughly deal with the states of qubit and give series of analytic coherence measures. The discussion and the conclusion are given in Sec. VI.

\section{The coherence monotone via pure-state coherence}
To begin with, let's give a brief introduction of the framework of the resource theory especially of coherence. The resource theory is well defined by the free state and the free operation \cite{qrt1,qrt2,qrt3}. For the coherence as a resource, the free state is the incoherent quantum states which can be given as $\delta=\sum_i \delta_i\left\vert i\right\rangle\left\langle i\right\vert$ with respect to the basis $\{\left\vert i\right\rangle\}$. The set of incoherent states is denoted by $\mathcal{I}$. The free operation (or the incoherent operation) is given by the completely-positive trace-preserving (CPTP) map defined in the Kraus representation as
$\varepsilon(\cdot)={\sum_nK_n(\cdot)K_n^\dag}$ with $K_n\delta K_n^\dag \in \mathcal{I}$ for an incoherent state $\delta$. Thus a good coherence measure $C(\rho)$  of a density matrix $\rho$  should satisfy the following conditions:  \cite{qc1} \\
($A_1$) {\textit {non-negativity}:} $C(\rho)\geq0$ is saturated iff $\rho\in \mathcal{I}$;\\
($A_2$) {\textit {monotonicity}:}  $C(\varepsilon(\rho))\leq C(\rho)$ for any incoherent operation $\varepsilon(\cdot)$;\\
($A_3$) {\textit {strong monotonicity}:}  $\sum_np_nC(K_n\rho K_n^{\dagger}/p_n)\leq C(\rho)$ with $p_n=\mathrm{Tr}[K_n\rho K_n^\dag]$ and $\rho_n=K_n\rho K_n^\dag/p_n$;\\
($A_4$) {\textit {convexity}:} $C(\rho)\leq \sum_ip_iC(\rho_i)$ for any $\rho=\sum_ip_i\rho_i$.\\
($A_5$) {\textit {only maximally coherent states (MCS) reach the maximum}:} $C(\rho)$ is maximal only for $\rho=|\Phi_d\rangle\langle\Phi_d|$ where $|\Phi_d\rangle=\frac{1}{\sqrt{d}}\sum_{n=1}^de^{i\theta_n}|n\rangle$ with real $\theta_n$ \cite{decom}.

In general,  $C(\cdot)$ is a good coherence measure if it satisfies all the above conditions. However,  $C(\cdot)$ will be called as a coherence monotone if it satisfies all the conditions but ($A_4$), which is similar to the entanglement monotone \cite{Plenio2014}. Here we would like to emphasize that a monotone is sometimes as important as a measure, since it is shown that  $C(\cdot)$ ( similar to the entanglement monotone) has its operational meaning  \cite{ne,ln} and in this sense, the convexity is usually understood as a mathematical convenience  \cite{entanglement,qrt1}.

In fact, the resource theory can always be established as long as a free state and a free operations are defined. Based on the different considerations for coherence, it has been shown that the free operations include at least five types \cite{free} such as physically incoherent operations (PIO)  defined by the operations  implemented only by incoherent unitary, incoherent ancillary system and incoherent projective measurement, maximally incoherent operations (MIO) defined by the operations that can convert one incoherent state to another incoherent state, dephasing-covariant incoherent operations (DIO) defined by the set of all maps commute with dephasing map,  incoherent operations (IO) which are defined as $\varepsilon(\cdot)$, and strictly incoherent operations (SIO)  defined as the subset of IO with the additional condition that  $\varepsilon^\dag(\cdot)$ is also IO.  Here we are mainly interested in  the IO and the SIO.

It isn't difficult to understand that a pure state can always be converted into a mixed state by some SIO/IO. On the contrary, for a mixed state, one can always find a pure state which can be converted into the given mixed state by SIO/IO (Note that SIO is a subset of IO). A typical example is that any mixed state can be considered as the pure state with the same diagonal entries as the given mixed state undergoes series of purely dephasing channels to reduce the moduli of the off-diagonal entries and undergoes some proper phase operations to adjust the phases. In particular, an incoherent mixed state can also correspond to an incoherent pure state in this sense. Note that the corresponding pure states for a given mixed state aren't generally unique. In
this sense, one can collect all these pure states as a set $R(\rho)$ corresponding to the certain mixed state $\rho$. In other words, \textit{the set  $R(\rho)$ isn't empty for any given state $\rho$}. Next, we will show that the coherence of the state $\rho$ can be well described by the minimal coherence achieved by the pure state $\left\vert \phi\right\rangle\in R(\rho)$.

To do so, we have to first consider a coherence measure of a pure state. Let $\overrightarrow{\mu}(\left\vert\psi\right\rangle)=\left(\left\vert\left\langle 1\right.\left\vert\psi\right\rangle\right\vert^2,\left\vert\left\langle 2\right.\left\vert\psi\right\rangle\right\vert^2,\cdots\right)$ denote the coherence vector with respect to the some basis $\{|i\rangle\}$. Denote $f(\overrightarrow{\mu})$ as a symmetric concave function with two additional conditions: (1) $f=0$ whenever $\overrightarrow{\mu}$ being a permutation of $(1,0,\cdots,0)$; (2) $f$ reaches the maximum only when every element of $\overrightarrow{\mu}$ equals to $1/d$ ($d$ is dimension of $\overrightarrow{\mu}$). It is shown that any good coherence measure can always be reduced to  a symmetric concave function $f(\overrightarrow{\mu})$ of $\overrightarrow{\mu}(\left\vert\psi\right\rangle)$ if applied on a pure state $\left\vert\psi\right\rangle$ \cite{oto1,oto2}. \textit{Throughout the paper,  we specify $F\left(\left\vert\psi\right\rangle\right)$ as a good pure-state coherence measure which means  $F\left(\left\vert\psi\right\rangle\right)$ is defined only for pure states by $f(\overrightarrow{\mu}(\left\vert\psi\right\rangle))$ mentioned above and satisfies ($A_1$)-($A_3$) and ($A_5$) for pure states}. In this sense,  $F\left(\left\vert\psi\right\rangle\right)$ doesn't pertain to mixed states, therefore the convexity given by ($A_4$) makes no sense. With the pure-state coherence measure $F\left(\left\vert\psi\right\rangle\right)$, we can further propose our coherence monotone for any mixed state in the following rigorous way.

{\bf{Theorem 1.-}} If $R(\rho)$ is the set of pure states that can be converted into the given state $\rho$ by IO, then $C_m(\rho)$ is a coherence monotone with
\begin{equation}
C_m(\rho)=\mathop {\inf}\limits_{\left\vert\phi\right\rangle\in R(\rho)}F(\left\vert\phi\right\rangle),\label{de}
\end{equation}
where $F\left(\left\vert\phi\right\rangle\right)$ is a good pure-state coherence measure mentioned above.

\textbf{Proof}:  In order to prove the theorem, we will have to show that $C_m(\cdot)$ satisfies all the conditions ($A_1$)-($A_3$) and ($A_5$).

($A_1$: \textit{Nonnegativity}) Suppose $\sigma=\sum_j\sigma_{jj}|j\rangle\langle j|$ is an arbitrary incoherent state, and $|1\rangle\langle 1|$ is an incoherent pure state. Define a SIO(IO) $\varepsilon_W=\{W_j\}$  as
\begin{align}
W_j&=\sum_{\gamma}b_{\gamma}^{(j)}|h_j(\gamma)\rangle\langle \gamma|,\\
&\sum_j\left\vert b_{\gamma}^{(j)}\right\vert^2=1,
\end{align}
where  $h_j$ is a permutation function with $h_j(\gamma)=\beta$ for the integers $\gamma$, $\beta$ and $h_j(\gamma_1)\ne h_j(\gamma_2)$ if $\gamma_1\ne\gamma_2$. Then,
\begin{align}
\varepsilon_W(|1\rangle\langle 1|)&=\sum_jW_j|1\rangle\langle 1|W_j^{\dagger}\nonumber\\
&=\sum_j\left\vert b_1^{(j)}\right\vert^2|h_j(1)\rangle\langle h_j(1)|.
\end{align}
If we let $h_j(1)=j$ and $|b_1^{(j)}|^2=\sigma_{jj}$, it is obvious that
\begin{align}
\varepsilon_W(|1\rangle\langle 1|)=\sigma,\label{inco}
\end{align}
which shows that for any incoherent state $\sigma$, one can always find a corresponding incoherent pure state $\left\vert 1\right\rangle$  such that $\left\vert 1\right\rangle$ can
be converted to $\sigma$ by SIO(IO). This implies that for any incoherent state $C_m=0$.
On the contrary, IO cannot convert an incoherent state to a coherent state, so for any coherent state $C_m>0$.

 ($A_2$: \textit{Monotonicity})
Let $\Lambda$ be an arbitrary IO and  $\rho$ denote any state. Suppose $|\psi\rangle\in R(\rho)$ is the optimal pure state subject to $C_m(\rho)=F(|\psi\rangle)$, then it is implied that $\varepsilon(|\psi\rangle\langle\psi|)=\rho$. Define $\rho_0=\Lambda(\rho)$, i.e., $\rho_0=\Lambda[\varepsilon(|\psi\rangle\langle\psi|)]$. Based on the definition of $C_m$ given in Eq.(\ref{de}), one can easily find $F(|\psi\rangle)\geq C_m(\rho_0)$, that is, $C_m(\rho)\geq C_m(\Lambda(\rho))$.

($A_3$: \textit{Strong monotonicity}) Let $\varepsilon_K(\cdot)=\sum_lK_l(\cdot)K_l^{\dagger}$ be an IO. For a state $\rho$, define
\begin{align}
p_l&=\mathrm{Tr}(K_l\rho K_l^{\dagger}),\nonumber\\
\rho_l&=K_l\rho K_l^{\dagger}/p_l.\label{tr1}
\end{align}
The strong monotonicity is equivalent to $C_m(\rho)\geq \sum_l p_l C_m(\rho_l)$.

Suppose $|\psi\rangle$ is the optimal state in $R(\rho)$ such that $C_m(\rho)=F(\left\vert \psi\right\rangle)$.
It is implied that the following relation holds:
\begin{align}
|\psi\rangle\stackrel{IO}{\longrightarrow}\rho\stackrel{\{K_l\}}{\longrightarrow}\{p_l,\rho_l\}.\label{stm}
\end{align}
Eq. (\ref{stm}) indicates that there exists an IO such that
\begin{align}
|\psi\rangle\stackrel{IO}{\longrightarrow}\{t_i,|\varphi_i\rangle\}\stackrel{\{K_l\}}{\longrightarrow}
\{t_iq_{il},|\phi_{il}\rangle\},
\end{align}
where $\rho=\sum_i t_i \left\vert \varphi_i\right\rangle\left\langle\varphi_i\right\vert$ with $t_i>0$ and $q_{il}=\mathrm{Tr}[K_l\left\vert\varphi_i\right\rangle\left\langle \varphi_i\right\vert K_l^\dag]$
and $|\phi_{il}\rangle=K_l\left\vert\varphi_i\right\rangle/\sqrt{q_{il}}$, $q_{il}\ne 0$. In other words, $|\psi\rangle$ can be converted into $\{t_iq_{il},|\phi_{il}\rangle\}$ by IO, which, based on Ref. \cite{tran2},  is equivalent to
\begin{align}
\mu^{\downarrow}(|\psi\rangle)\prec\sum_{i,l}t_iq_{il}\mu^{\downarrow}(|\phi_{il}\rangle),\label{prove1}
\end{align}
where $\mu^{\downarrow}(|\psi\rangle)$ is the coherence vector in decreasing order. Define the pure state $|\psi_l\rangle$ such that
\begin{align}
\mu^{\downarrow}(|\psi_l\rangle)=\sum_i\frac{t_iq_{il}}{p_l}\mu^{\downarrow}(|\phi_{il}\rangle).\label{prove2}
\end{align}
It's obvious $p_l=\sum_it_iq_{il}$, $\rho_l=\sum_i\frac{t_iq_{il}}{p_l}|\phi_{il}\rangle\langle\phi_{il}|$. One can directly arrive at $\mu^{\downarrow}(|\psi_l\rangle)\prec\sum_i\frac{t_iq_{il}}{p_l}\mu^{\downarrow}(|\phi_{il}\rangle)$ which, based on Ref. \cite{tran2}, shows that
 $|\psi_l\rangle$ can be converted into $\rho_l$ by IO. According to the definition of $C_m$, we have
\begin{align}
F(|\psi_l\rangle)\geq C_m(\rho_l).\label{prove3}
\end{align}
Substituting Eq. (\ref{prove2}) into Eq. (\ref{prove1}), one can obtain
\begin{align}
\mu^{\downarrow}(|\psi\rangle)\prec&\sum_lp_l\sum_i\frac{t_iq_{il}}{p_l}\mu^{\downarrow}(|\phi_{il}\rangle)\nonumber\\
=&\sum_lp_l\mu^{\downarrow}(|\psi_l\rangle).
\end{align}
It states that $|\psi\rangle$ can be converted to $\{p_l,|\psi_{l}\rangle\}$ by IO, so the strong monotonicity of the selected measure $F(\cdot)$ gives
\begin{align}
F(|\psi\rangle)\geq\sum_lp_lF(|\psi_{l}\rangle),\label{one}
\end{align}
Substituting  Eq. (\ref{prove3}) into Eq.(\ref{one}), one can obtain
\begin{align}
C_m(\rho)=F(|\psi\rangle)\geq&\sum_lp_lF(|\psi_l\rangle)\nonumber\\
\geq&\sum_lp_lC_m(\rho_l),
\end{align}
which is the exact strong monotonicity of $C_m$.

($A_5$:  \textit{Only MCS reach the maximum.}) Suppose that $\rho=\sum_ip_i|\varphi_i\rangle\langle\varphi_i|$ isn't the MCS. Ref. \cite{decom} shows that there is at least one pure state $|\varphi_{i_0}\rangle$ which isn't the MCS.  Thus $\mu^{\downarrow}(|\Phi_d\rangle)\prec\sum_ip_i\mu^{\downarrow}(|\varphi_i\rangle)$  and $\mu^{\downarrow}(|\Phi_d\rangle)\ne\sum_ip_i\mu^{\downarrow}(|\varphi_i\rangle)$, which implies $\left\vert \Phi_d\right\rangle$ can be converted into $\rho$. Define $|\psi\rangle$ such that $\mu^{\downarrow}(|\psi\rangle)=\sum_ip_i\mu^{\downarrow}(|\varphi_i\rangle)$. One will obtain that  $|\psi\rangle \neq |\Phi_d\rangle$, $\left\vert \psi\right\rangle$ can be converted to $\rho$, and $\left\vert \Phi_d\right\rangle$ can be converted into $\left\vert \psi\right\rangle$. Based on the monotonicity of $F$, one can see that  $F(|\Phi_d\rangle)>F(|\psi\rangle)\geq F(\rho)$. Based on the definition of  $C_m$, one can find that  $F(|\Phi_d\rangle)>F(|\psi\rangle)\geq C_m(\rho)$, which shows any state $\rho$ which isn't the MCS cannot reach maximum. Conversely, from Eq.(\ref{de}), $C_m$ inherits property ($A_5$) of $F$ for pure states.  $\hfill\blacksquare$

With the above theorem, next we will show that our proposed coherence monotone $C_m(\rho)$ serves as the supremum of all the coherence monotones which reduced to $F$ for pure states.

\textbf{Corollary 1}.- For any coherence monotone $C(\cdot)$  with $C(\left\vert \psi\right\rangle)=C_m(\left\vert\psi\right\rangle)$ for any pure state $\left\vert\psi\right\rangle$, $C_m(\rho)\geq C(\rho)$ holds for any state $\rho$.

\textbf{Proof.}  Given a density matrix $\rho$, based on the definition of $C_m(\rho)$, one can always find the corresponding optimal pure state $\left\vert \psi\right\rangle$ such that $C_m(\rho)=F(\left\vert\psi\right\rangle)$ with $\rho$ obtained by IO on the optimal pure state $\left\vert \psi\right\rangle$.
Note that it is also valid to write $C_m(\rho)=C_m(\left\vert\psi\right\rangle)=C(\left\vert\psi\right\rangle)=F(\left\vert\psi\right\rangle)$. Since $C(\cdot)$  is also a coherence monotone, we have $C(\left\vert\psi\right\rangle)\geq C(\rho)$, which implies $C_m(\rho)\geq C(\rho)$. The proof is completed.$\hfill\blacksquare$

Up to now, a valid coherence monotone $C_m(\cdot)$ has been completely established if a pure-state coherence measure $F(\cdot)$ is given.
Based on our definition of $C_m(\cdot)$, one can see that $C_m(\rho)$ of the state $\rho$ is obtained by the minimal pure-state coherence optimized in the set $R(\rho)$. We will show that the set $R(\rho)$ in the above minimization can be actually replaced by its  subset denoted by $Q(\rho)$. So our coherence measure $C_m(\rho)$ can be rewritten based on $Q(\rho)$. For clarity, we'd like to give the explicit forms of both $Q(\rho)$ and  $C_m(\rho)$ in the following rigorous way.

\textbf{Theorem 2}.-The coherence  monotone $C_m(\rho)$ of a density matrix $\rho$ can be rewritten as
\begin{equation}
C_m(\rho)=\inf_{\left\vert\psi\right\rangle\in Q(\rho)} F(\left\vert \psi\right\rangle),
\end{equation}
where $F(\left\vert\psi\right\rangle)$ is defined the same as Theorem 1, and $Q(\rho)\subset R(\rho)$ is the set of all pure states $\left\vert \phi\right\rangle$ which fulfill
\begin{equation}
\mu^{\downarrow}(|\phi\rangle)=\sum_ip_i\mu^{\downarrow}(|\varphi_i\rangle),
\end{equation}
where $\{p_i,\left\vert \varphi_i\right\rangle\}$ is a pure-state decomposition of $\rho$.

\textbf{Proof}.-Let $|\psi\rangle\in R(\rho)$, then there exists a decomposition $\{p_i,|\varphi_i\rangle\}$ of $\rho$ such that  $\mu^{\downarrow}(|\psi\rangle)\prec\sum_ip_i\mu^{\downarrow}(|\varphi_i\rangle)$ \cite{tran2}. Define a pure state $\left\vert\psi_0\right\rangle$ such that $\mu^{\downarrow}(|\psi_0\rangle)=\sum_ip_i\mu^{\downarrow}(|\varphi_i\rangle)$ which actually implies $\mu^{\downarrow}(|\psi_0\rangle)\prec\sum_ip_i\mu^{\downarrow}(|\varphi_i\rangle)$ and $\mu^{\downarrow}(|\psi\rangle)\prec\mu^{\downarrow}(|\psi_0\rangle)$, then we have that $\left\vert \psi\right\rangle$ can be converted into $\left\vert\psi_0\right\rangle$ and $\left\vert \psi_0\right\rangle$ can be converted to $\rho$. Correspondingly, it follows that $F(\left\vert\psi_0\right\rangle)\leq F(\left\vert\psi\right\rangle)$. Thus all the pure states $|\psi_0\rangle$ can form the subset  $Q(\rho)$. In particular, one can find that the minimal $F(\cdot)$ can be achieved by  those in the subset  $Q(\rho)$.$\hfill\blacksquare$

\section{The convexity}
In the previous section, we don't address the convexity. Now we will study the requirements of $F(\cdot)$ such that our proposed coherence monotone can become a good coherence measure, that is, $C_m(\cdot)$ is convex.

\textbf{Theorem 3}.-$C_m$ is convex if and only if for any ensemble $\{p_i,|\psi_i\rangle\}$ (let $\varrho=\sum_ip_i|\psi_i\rangle\langle\psi_i|$), there always exists a pure state $|\varphi_0\rangle\in R(\varrho)$ such that
\begin{align}
F(|\varphi_0\rangle)\leq\sum_ip_i F(|\psi_i\rangle).\label{co1}
\end{align}

\textbf{Proof}.-Suppose $\{p_i,|\psi_i\rangle\}$ is an arbitrary ensemble and $\varrho=\sum_ip_i|\psi_i\rangle\langle\psi_i|$. If there exists  $|\varphi_0\rangle\in R(\varrho)$ satisfying Eq.(\ref{co1}), then
\begin{align}
C_m(\varrho)\leq F(|\varphi_0\rangle)\leq\sum_ip_iF(|\psi_i\rangle)=\sum_ip_iC_m(|\psi_i\rangle).\label{E1}
\end{align}
Corollary 1 shows that $C_m$ is the upper bound of any coherence monotone which gives the same coherence as $C_m$ for pure states, hence $C_m$ is not less than $C_f$, the coherence measure based on the convex roof construction, i.e.
\begin{align}
C_m(\varrho)\geq C_f(\varrho)=\inf\limits_{\{t_l,|\chi_l\rangle\}} \sum_lt_lC_m(|\chi_l\rangle)\label{E2}
\end{align}
with $\varrho=\sum_{l} t_l\left\vert \chi_l\right\rangle\left\langle \chi_l\right\vert$ and $\sum_l t_l=1,t_l>0$. If $\{p_i,\left\vert \psi_i\right\rangle\}$  in Eq. (\ref{E1}) happens to be the optimal
decomposition that achieves $C_f(\varrho)$ in Eq. (\ref{E2}), one can easily obtain that $C_m(\varrho)=C_f(\varrho)$. It implies that $C_m(\varrho)$ inherits the convexity of $C_f(\varrho)$.

Conversely, let $\varrho=\sum_ip_i|\psi_i\rangle\langle\psi_i|$ and $|\varphi_0\rangle\in R(\varrho)$ be the optimal state such that $F(|\varphi_0\rangle)=C_m(\varrho)$.
If $C_m$ is convex, then
\begin{align}
F(|\varphi_0\rangle)=C_m(\varrho)\leq\sum_ip_iF(|\psi_i\rangle).
\end{align}
The proof is completed.$\hfill\blacksquare$

Theorem 3 shows that if the conditions Eq. (\ref{co1}) are satisfied, the proposed coherence measure $C_m$ is a good coherence measure. In fact, if $C_m$ is convex,
$C_m$ can own more general important properties.

\textbf{Theorem 4}.-For a state $\rho$, $C_m(\rho)=C_f(\rho)$ is equivalent to that $C_m$ is convex, where $C_f(\rho)$ is the coherence measure in terms of the convex roof construction.

\textbf{Proof}. The proof actually is given in the proof of Theorem 3, so it isn't repeated here. $\hfill\blacksquare$

As mentioned at the beginning of the last section, the main results are only restricted to the case of IO/SIO, so the coherence strong monotone can be established first and then in the current section we mainly consider the convexity. However,  if $C_m$ satisfying the convexity is a prerequisite, one will find from the following thereom that our approach is also suitable for the establishment of the coherence measure in the sense of MIO, DIO and PIO.

\textbf{Theorem 5}.-If  $C_m(\cdot)$ is convex, and $F(\cdot)$ for pure states satisfies the strong monotonicity with respect to MIO (DIO, IO, PIO or SIO), then for any state $\rho$, $C_m(\rho)$ will also satisfies the strong monotonicity with respect to MIO ( DIO, IO, SIO or PIO).

\textbf{Proof}.-For a certain $\rho$, let $\left\vert\psi\right\rangle$ be the optimal pure state such that $F(\left\vert \psi\right\rangle)=C_m(\rho)$ with $\rho=\varepsilon_M(\psi)$ ($\varepsilon_M$ is an MIO). Suppose that $\varepsilon_K$ is an arbitrary MIO, it's clear that $\varepsilon_T(\cdot)=\varepsilon_K\circ\varepsilon_M(\cdot)$ must be an MIO.  Let the Kraus operators of $\varepsilon_K$, $\varepsilon_M$ and $\varepsilon_T$ be denoted  respectively by $\{K_i\}$, $\{M_l\}$ and $\{T_{il}\}$ with $q_{il}=\left\langle \psi\right\vert T_{il}^\dag T_{il}\left\vert \psi\right\rangle$ and $p_i=\left\langle \psi\right\vert K_i^\dag K_i\left\vert \psi\right\rangle$. Then
\begin{align}
C_m(\rho)&=F(\left\vert\psi\right\rangle)\nonumber\\
&\geq \sum_{il}q_{il}F(T_{il}|\psi\rangle\langle\psi| T_{il}^{\dagger}/q_{il})\nonumber\\
&=\sum_ip_i\sum_l\frac{q_{il}}{p_i}F(T_{il}\left\vert \psi\right\rangle\left\langle \psi\right\vert T_{il}^{\dagger}/q_{il})\nonumber\\
&\geq\sum_ip_iC_m(\sum_l\frac{q_{il}}{p_i}K_iM_l\left\vert \psi\right\rangle\left\langle \psi\right\vert M_l^{\dagger}K_i^{\dagger}/q_{il})\nonumber\\
&=\sum_ip_iC_m(K_i\rho K_i^{\dagger}/p_i),\label{io}
\end{align}
where the first inequality is due to strong monotonicity of $F(\cdot)$ under MIO and the second inequality is due to the convexity of $C_m(\cdot)$. It is especially noted that when $\varepsilon_K$ is a DIO, IO, SIO or PIO, $\varepsilon_T$ will also be DIO, IO, SIO or PIO. Thus the same proof also holds, so if for pure states $F(\cdot)$ satisfies the strong monotonicity under MIO (DIO, IO, SIO or PIO)  and $C_m$ is convex, $C_m$ also satisfies the strong monotonicity under MIO (DIO, IO, SIO or PIO).$\hfill\blacksquare$

\section{Examples}
\subsection{The geometric coherence as $F$}
In the previous sections, we have given the general form of our new coherence measure. Next, we will give a concrete example by selecting an exact coherence measure $F(\cdot)$ for pure states. Here we'd like to choose the geometric coherence $C_g$ as the candidate which defined as  \cite{op2}
\begin{align}
C_g(|\psi\rangle)=1-\sup\limits_{\sigma\in \mathcal{I}}\mathcal{F}(|\psi\rangle,\sigma),\label{g1}
\end{align}
where $\mathcal{F}(|\psi\rangle,\sigma)={\left\langle\psi\right\vert\sigma\left\vert \psi\right\rangle}$ is the fidelity between the pure state $\left\vert\psi\right\rangle$ and the state $\sigma$, then we have the following theorem.

{\textbf{Theorem 6}}.- The coherence of a state $\rho$ can be well measured by
\begin{align}
C_m^g(\rho)=\inf\limits_{|\phi\rangle\in Q(\rho)}C_g(|\phi\rangle)=\inf_{\{p_i,\left\vert\psi_i\right\rangle\}}\sum p_i C_g(\left\vert\psi_i\right\rangle)
\end{align}
 with $\rho=\sum p_i\left\vert \psi_i\right\rangle\left\langle \psi_i\right\vert$.

\textbf{Proof}.- It is obvious that the geometric coherence per se is a coherence monotone, so the key task is to prove $C_m^g$ is convex. However, we don't
directly show that the geometric coherence $C_g$ satisfies Theorem 3, but we will prove that $C_m^g$ is actually a coherence measure based on the convex roof construction which
will imply that $C_m^g$ is a good coherence measure (especially satisfies the convexity).

Without loss of generality, let's consider the coherence in the framework defined by  the computational basis $\{\left\vert i\right\rangle\}$. It is obvious that  for a pure state $\left\vert\varphi\right\rangle$, we have
\begin{align}
C_g(|\varphi\rangle)=&1-\sup\limits_{i}|\langle\varphi|i\rangle|^2\nonumber\\
=&1-\mu^{\downarrow}(|\varphi\rangle)_1
\end{align}
with $\mu^{\downarrow}(|\varphi\rangle)_1$ denoting the first element of $\mu^{\downarrow}(|\varphi\rangle)$. Now we take the geometric coherence $C_g(\rho)$ as the pure-state coherence measure $F(\cdot)$, then
\begin{align}
C_m^g(\rho)=&\inf\limits_{|\varphi\rangle\in Q(\rho)}C_g(|\varphi\rangle)\nonumber\\
=&1-\sup\limits_{|\varphi\rangle\in Q(\rho)}\mu^{\downarrow}(|\varphi\rangle)_1.
\end{align}
Based on Theorem 5, one can note that $|\varphi\rangle\in Q(\rho)$ means that there exists a decomposition $\{p_i,|\psi_i\rangle\}$ of $\rho$ such that
\begin{align}
\mu^{\downarrow}(|\varphi\rangle)_1=\sum_ip_i\mu^{\downarrow}(|\psi_i\rangle)_1.
\end{align}
Thus, $C_m^g(\rho)$ can be rewritten as
\begin{align}
C_m^g(\rho)=&1-\sup\limits_{\{p_i,\left\vert\psi_i\right\rangle\}}\sum_ip_i\mu^{\downarrow}(|\psi_i\rangle)_1\nonumber\\
=&\inf\limits_{\{p_i,\left\vert\psi_i\right\rangle\}}p_iC_g(|\psi_i\rangle),
\end{align} which shows that $C_m^g$ is the coherence measure based on the convex roof construction. So it automatically satisfies the convexity. $\hfill\blacksquare$

\subsection{Analytical expressions for qubits}

Now we will study the potential analytic expression of our proposed coherence measure.
Based on our definition, one can easily note that the key of calculating our coherence measure is whether one could find out the optimal pure state $\left\vert \psi\right\rangle\in Q(\rho)$
such that $C_m(\rho)=F(\left\vert \psi\right\rangle)$.
First we would like to give the following lemma.

\textbf{Theorem 7}.-If there exists an optimal decomposition $\{\tilde{p}_j,|\tilde{\phi}_j\rangle\}$ for the state $\rho$ such that
\begin{align}
\sum_ip_i\mu^{\downarrow}(|\phi_i\rangle)\prec\sum_j\tilde{p}_j\mu^{\downarrow}(|\tilde{\phi}_j\rangle) \label{youyong}
\end{align}
with  $\{p_i,|\phi_i\rangle\}$ denoting any decomposition of $\rho$, the optimal pure state $\left\vert \psi\right\rangle$ can be defined by
 \begin{equation}\mu^{\downarrow}(|\psi\rangle)=\sum_j\tilde{p}_j\mu^{\downarrow}(|\tilde{\phi}_j\rangle).\label{31}\end{equation}

\textbf{Proof}. To show this, let's suppose there exists the optimal decomposition $\{\tilde{p}_j,|\tilde{\phi}_j\rangle\}$ of $\sigma$ subject to Eq. (\ref{youyong}). Thus we can  denote $\mu^{\downarrow}(|\psi\rangle)=\sum_j\tilde{p}_j\mu^{\downarrow}(|\tilde{\phi}_j\rangle)$. Considering any state $|\varphi\rangle\in Q(\rho)$,
there always exists a decomposition $\{p_i,|\phi_i\rangle\}$ such that
\begin{align}
\mu^{\downarrow}(|\varphi\rangle)=&\sum_ip_i\mu^{\downarrow}(|\phi_i\rangle)
\prec\sum_j\tilde{p}_j\mu^{\downarrow}(|\tilde{\phi}_j\rangle)
=\mu^{\downarrow}(|\psi\rangle).\label{dede}
\end{align}
This shows that $|\varphi\rangle$ can be converted into $\left\vert \psi\right\rangle$ by IO, that is, $F(|\varphi\rangle)\geq F(\left\vert \psi\right\rangle)$.
In other words, $\left\vert \psi\right\rangle$ can achieve the least coherence of the pure state $|\varphi\rangle\in Q(\rho)$, namely, Eq. (\ref{31}) holds.$\hfill\blacksquare$

One can find that to obtain the analytic expression, whether there exists a decomposition as Eq. (\ref{youyong}) is the key. However, it is not easy to prove whether there always exists such an optimal decomposition for a general quantum state $\rho$. But we can show that such an optimal decomposition can always be found in qubit states. That is, one
 can always establish the analytic coherence measure.

\textbf{Theorem 8}.-  Given a density matrix $\sigma$ of a qubit with $b$ denoting its off-diagonal element, the optimal decomposition subject to Eq. (\ref{youyong}) can be given by
 \begin{align}
\sigma=\lambda\sigma^{(+)}+(1-\lambda)\sigma^{(-)},\label{decomposition}
\end{align}
where $\sigma^{(\pm)}=\left(
\begin{array}{cc}
\frac{1\pm z}{2} & b \\
b^* & \frac{1\mp z}{2}
\end{array}\right)$ with $z=\sqrt{1-4|b|^2}$, and  $\lambda\in[0,1]$  is some weight parameter determined by the state $\sigma$. In this sense, the coherence can be given by
$C_m(\rho)=F(\sigma^{\pm}).$

\textbf{Proof}.-Suppose that $\{p_i,|\psi_i\rangle\}$ is any decomposition of $\sigma$, then similar to the state $\sigma^{\pm}$, we can use $\{b,z,\lambda \}$, $\{b_i,z_i\}$ to
express the states $\sigma$ and $\left\vert\psi_i\right\rangle\left\langle\psi_i\right\vert$ with $z=\sqrt{1-4|b|^2},z_i=\sqrt{1-4|b_i|^2}$. Considering $\rho=\sum_i p_i\left\vert\psi_i\right\rangle\left\langle\psi_i\right\vert$, we have
\begin{align}
z=&\sqrt{1-4|b|^2}=\sqrt{1-4|\sum_ip_ib_i|^2}\nonumber\\
\geq&\sqrt{1-4(\sum_ip_i|b_i|)^2}\geq\sum_ip_i\sqrt{1-4|b_i|^2}\nonumber\\
=&\sum_ip_iz_i.\label{34}
\end{align}
According to the definition of the coherence vector $\mu^{\downarrow}$ for pure states, from Eq. (\ref{34}) one can obtain
\begin{align}
\lambda\mu^{\downarrow}(\sigma^{(+)})+(1-\lambda)\mu^{\downarrow}(\sigma^{(-)})=(\frac{1+z}{2},\frac{1-z}{2})\nonumber\\
\succ(\frac{1+\sum_i p_i z_i}{2},\frac{1-\sum_i p_i z_i}{2})=\sum_ip_i\mu^{\downarrow}(|\psi_i\rangle),
\end{align}
which proves the existence of the required optimal decomposition. In addition, based on Eq. (\ref{dede}), one can know that
$\sigma^{\pm}$ is the exact optimal pure state in $Q(\rho)$ such that $C_m(\rho)=F(\sigma^{\pm})$. The proof is completed. $\hfill\blacksquare$

\textbf{Theorem 9}.-For a qubit density matrix, the conditions for convexity given in Theorem 3 is
equivalent to that  $F(|\varphi\rangle)=f(|b|)$  is a convex function on $\left\vert b\right\vert$ for the pure state $|\varphi\rangle\langle\varphi|=\left(
\begin{array}{cc}
\frac{1\pm z}{2} & b \\
b^* & \frac{1\mp z}{2}
\end{array}\right)$ with $z=\sqrt{1-4\left\vert b\right\vert^2}$.

\textbf{Proof}.-Any pure state of qubit can be written as the form of  $\left\vert \varphi\right\rangle$, so its coherence vector can be given as
$\mu^{\downarrow}(\left\vert \varphi\right\rangle)=(\frac{1+z}{2},\frac{1-z}{2})$. Similarly, for another pure state $\left\vert\psi\right\rangle$, we can denote its coherence vector as
$\mu^{\downarrow}(\left\vert \psi\right\rangle)=(\frac{1+z'}{2},\frac{1-z'}{2})$. For a good coherence monotone $F$,
one has $F(\left\vert \psi\right\rangle)\leq F(\left\vert \varphi\right\rangle)$ if $\mu^{\downarrow}(\left\vert \varphi\right\rangle)\prec\mu^{\downarrow}(\left\vert \psi\right\rangle)$
 which implies that $\left\vert \varphi\right\rangle$ can be converted to $\left\vert \psi\right\rangle$ by IO. Thus we can easily find that $z'\geq z$, which indicates that $F(\left\vert \varphi\right\rangle)$ is a monotonically decreasing function on $z$. Since $z=\sqrt{1-4\left\vert b\right\vert^2}$, we can equivalently
  say that $F(\left\vert \varphi\right\rangle)=f(\left\vert b\right\vert)$ is a monotonically increasing function on $\left\vert b\right\vert$.

To prove $f$ is a convex function, we consider a particular state $\sigma=\sum_ip_i|\psi_i\rangle\langle\psi_i|$ where we denote the off-diagonal entries of $|\psi_i\rangle\langle\psi_i|$ by $|b_i|$.
So the off-diagonal entry $b$ of $\sigma$ can be written as $|b|=b=\sum_ip_i|b_i|$. Let $|\varphi_0\rangle\in Q(\sigma)$ and $\{p_i,|\psi_i\rangle\}$ be the exact pure state and the corresponding decomposition of $\sigma$ required in Theorem 3, then Theorem 3 shows
\begin{align}
F(|\varphi_0\rangle)\leq\sum_ip_iF(|\psi_i\rangle).
\end{align}
Let $|\varphi\rangle$ is the optimal state such that  $C_m(\sigma)=F(|\varphi\rangle)$, with $C_m(\sigma)\leq F(|\varphi_0\rangle)$ due to $|\varphi_0\rangle\in Q(\sigma)$,
one will immediately find
\begin{align}
F(|\varphi\rangle)\leq\sum_ip_iF(|\psi_i\rangle).
\end{align}
Theorem  8 implies that the off-diagonal element of $|\varphi\rangle\langle\varphi|$ can be the same as $\sigma$, so we can write
\begin{align}
f(\left\vert b\right\vert)=F(|\varphi\rangle)\leq\sum_ip_iF(|\psi_i\rangle)=\sum_ip_if(\left\vert b_i\right\vert),
\end{align}
which shows the convex $f$.

Conversely, we first assume $f$ is convex.  Let $|\phi\rangle$ be the optimal state in $Q(\sigma)$ such that $F(|\phi\rangle)=C_m(\sigma)$, then Theorem 8 shows that $\left\vert\phi\right\rangle\left\langle\phi\right\vert$ can have the same off-diagonal entry $b$ as $\sigma$. Suppose $\sigma=\sum_i p_i\left\vert \psi_i\right\rangle\left\langle\psi_i\right\vert$ with $b_i$ denoting the off- diagonal  entries of $\left\vert \psi_i\right\rangle\left\langle\psi_i\right\vert$, then we have
\begin{align}
F(|\phi\rangle)=&f(|b|)=f(|\sum_ip_ib_i|)\nonumber\\
\leq& f(\sum_ip_i|b_i|)\leq\sum_ip_if(|b_i|)\nonumber\\
=&\sum_ip_iF(|\psi_i\rangle),\label{rr}
\end{align}
where the first inequality comes from the monotonically increasing function $f$ on $|b|$, and the second inequality is attributed to the convexity.  Eq. (\ref{rr}) is exactly the same as Eq. (\ref{co1}). The proof is completed.$\hfill\blacksquare$

Based on the above theorems, we have known the optimal pure state $\left\vert\varphi\right\rangle$ for a mixed state $\sigma$ of qubit. So one can easily select the coherence measure $F$ for pure state, then use $F$ to measure the coherence of $\left\vert\varphi\right\rangle$ and finally obtain the coherence $C_m(\sigma)=F(\left\vert\varphi\right\rangle)$. Here we would like to emphasize that almost all the known coherence measures based on $l_1$ norm, relative entropy, geometric coherence, skew information and so on are convex on $|b|$ for a pure state of qubit. Therefore, all these measures can be safely employed for $F$. The concrete expressions are omitted just because they become trivially simple due to our theorems.

\section{Discussion and Conclusions}

In conclusion, we have presented a new approach to quantifying quantum coherence. Our coherence measure can be understood as the least coherence of the pure states which can be converted to the state of interest. In particular, we have shown that our coherence monotone is the supremum of all the coherence monotones that have the same coherence for any given pure state. Our coherence measure is proven to be a subset of the coherence measure in terms of the convex roof construction, which gives a new understanding of the coherence measure. As the demonstration, we give the concrete example for our coherence measure. It is especially important that we have thoroughly analyze the case of qubit states and give series of analytic expressions of coherence. In addition, the same understanding approach could  also be  suitable for other resource theories, which will be studied in the forthcoming work.
\section*{Acknowledgements}
This work was supported by the
National Natural Science Foundation of China, under Grant No.11775040 and
No. 11375036, and the Fundamental
Research Fund for the Central Universities under Grants No. DUT18LK45.
\appendix
\section{An alternative proof of strong monotonicity subject to SIO}
{\bf{Lemma 3}}.-
Given a SIO operated on the state $\rho$ as $\varepsilon_K(\rho)=\sum_iK_i\rho K_i^{\dagger}$, there always exists a corresponding SIO on the pure state $\left\vert\psi\right\rangle\in R(\rho)$ as
$\varepsilon_T(\left\vert\psi\right\rangle\left\langle\psi\right\vert)=\sum_iT_i\left\vert\psi\right\rangle\left\langle\psi\right\vert T_i^{\dagger}$ such that
\begin{equation}
\mathrm{Tr}(T_i|\psi\rangle\langle\psi| T_i^{\dagger})=\mathrm{Tr}(K_i\rho K_i^{\dagger})\label{tr}
\end{equation}
for any given $i$. In addition,
The state $T_i|\psi\rangle\langle\psi| T_i^{\dagger}/\mathrm{Tr}(T_i|\psi\rangle\langle\psi| T_i^{\dagger})$ can be converted  into $K_i\rho K_i^{\dagger}/\mathrm{Tr}(K_i\rho K_i^{\dagger})$ by SIO.

\textbf{Proof}.-
For an $n$-dimensional pure state $|\psi\rangle$ given, with respect to the computational basis, by
\begin{equation}
 |\psi\rangle\langle\psi|=\sum_m|c_m|^2|m\rangle\langle m|+\sum_{m\ne n}c_mc_n^*|m\rangle\langle n|,
\end{equation}
let's consider three SIO  $\varepsilon_M=\{M_l\}$, $\varepsilon_K=\{K_i\}$ and $\varepsilon_T=\{T_j\}$ defined, respectively, by
\begin{equation}
M_l=\sum_{\gamma}a_{\gamma}^{(l)}|\pi_l(\gamma)\rangle\langle \gamma|,l=1,2,\cdots,
\end{equation}
\begin{equation}
K_i=\sum_{\gamma}\tau_{\gamma}^{(i)}|f_i(\gamma)\rangle\langle \gamma|,i=1,2,\cdots,
\end{equation}
and
\begin{align}
T_j=\sum_{\gamma}d_\gamma^{(j)}|\gamma\rangle\langle \gamma|,
\end{align}
where $\{\left\vert\gamma\right\rangle\}$ is the computational basis, $\pi_l$ and $f_i$ are the permutation operation labeled by $l$ and $i$ respectively,   $\sum_l\left\vert a_{\gamma}^{(l)}\right\vert^2=\sum_i\left\vert\tau_{\gamma}^{(i)}\right\vert^2=1$, and
 \begin{equation}\left\vert d_\gamma^{(i)}\right\vert^2=\sum_l\left\vert a_{\gamma}^{(l)}\right\vert^2\left\vert\tau_{\pi_l(\gamma)}^{(i)}\right\vert^2.\label{equ}\end{equation} Thus one can easily find that  $\sum_i\left\vert d_\gamma^{(i)}\right\vert^2=\sum_l\left\vert a_{\gamma}^{(l)}\right\vert^2\sum_i\left\vert\tau_{\pi_l(\gamma)}^{(i)}\right\vert^2=1$. With the above SIO, we have
\begin{align}
&K_iM_l|\psi\rangle\langle\psi| M_l^{\dagger}K_i^\dag\nonumber\\
=&\sum_m|c_m|^2\left\vert a_m^{(l)}\right\vert^2K_i|\pi_l(m)\rangle\langle\pi_l(m)|K_i^\dag\nonumber\\
+&\sum_{m\ne n}c_mc_n^*a_m^{(l)}a_n^{(l)*}K_i|\pi_l(m)\rangle\langle\pi_l(n)|K_i^\dag\nonumber\\
=&\sum_m|c_m|^2\left\vert a_m^{(l)}\right\vert^2\left\vert\tau^{(i)}_{\pi_l(m)}\right\vert^2|\pi_l(m)\rangle\langle\pi_l(m)|\nonumber\\
+&\sum_{m\ne n}c_mc_n^*a_m^{(l)}a_n^{(l)*}\tau^{(i)}_{\pi_l(m)}\tau^{(i)*}_{\pi_l(n)}|f_i[\pi_l(m)]\rangle\langle f_i[\pi_l(n)]|.
\end{align}
Therefore,
\begin{align}
&\mathrm{Tr}(K_i\rho K_i^{\dagger})=\sum_l\mathrm{Tr}(K_iM_l|\psi\rangle\langle\psi| M_l^{\dagger}K_i^{\dagger})\nonumber\\
=&\sum_l\sum_m|c_m|^2\left\vert a_m^{(l)}\right\vert^2\left\vert\tau_{\pi_l(m)}^{(i)}\right\vert^2.\label{s1}
\end{align}
Similarly,
\begin{align}
&\mathrm{Tr}[T_i|\psi\rangle\langle\psi| T_i^{\dagger}]\nonumber\\
=&\sum_m|c_m|^2\left\vert d_m^{(i)}\right\vert^2\left\langle m\right.\left\vert m\right\rangle
+\sum_{m\ne n}c_mc_n^*d_m^{(i)}d_n^{(i)*}\left\langle n\right\vert \left.m\right\rangle\nonumber\\
=&\sum_m|c_m|^2\left\vert d_m^{(i)}\right\vert^2.\label{s2}
\end{align}
Based on Eq. (\ref{equ}), it is obvious that Eqs. (\ref{s1}) and (\ref{s2}) imply $\mathrm{Tr}(K_i\rho K_i^{\dagger})=\mathrm{Tr}[T_i|\psi\rangle\langle\psi| T_i^{\dagger}]$. Thus, Eq. (\ref{tr}) is proved.

To proceed, let's consider an SIO $\varepsilon_N^{(i)}=\{N_l\}$ defined by
\begin{align}
N_l=\sum_\gamma\frac{a_\gamma^{(l)}\tau_{\pi_l(\gamma)}^{(i)}}{d_\gamma^{(i)}}|f_i[\pi_l(\gamma)]\rangle\langle \gamma|
\end{align}
with
\begin{align}
&\sum_lN_l^{\dagger}N_l\nonumber\\
=&\sum_l\sum_\gamma\frac{\left\vert a_\gamma^{(l)}\right\vert^2\left\vert\tau_{\pi_l(\gamma)}^{(i)}\right\vert^2}{\left\vert d_\gamma^{(i)}\right\vert^2}|\gamma\rangle\langle \gamma|\nonumber\\
=&\sum_\gamma|\gamma\rangle\langle \gamma|\frac{\sum_l\left\vert a_\gamma^{(l)}\right\vert^2\left\vert\tau_{\pi_l(\gamma)}^{(i)}\right\vert^2}{\left\vert d_\gamma^{(i)}\right\vert^2}\nonumber\\
=&\sum_\gamma|\gamma\rangle\langle \gamma|.
\end{align}
Then one can see that
\begin{align}
&\varepsilon_N(T_i|\psi\rangle\langle\psi| T_i^{\dagger})=\sum_lN_lT_i|\psi\rangle\langle\psi| T_i^{\dagger}N_l^{\dagger}\nonumber\\
=&\sum_lN_l(\sum_m|c_m|^2\left\vert d_m^{(i)}\right\vert^2|m\rangle\langle m|+\sum_{m\ne n}c_mc_n^*d_m^{(i)}d_n^{(i)*}|m\rangle\langle n|)N_l^{\dagger}\nonumber\\
=&\sum_l\sum_m|c_m|^2\left\vert a_m^{(l)}\right\vert^2\left\vert\tau_{\pi_l(m)}^{(i)}\right\vert^2|f_i[\pi_l(m)]\rangle\langle f_i[\pi_l(m)]|\nonumber\\
+&\sum_l\sum_{m\ne n}c_mc_n^*a_m^{(l)}a_n^{(l)*}\tau_{\pi_l(m)}^{(i)}\tau_{\pi_l(n)}^{(i)*}
|f_i[\pi_l(m)]\rangle\langle f_i[\pi_l(n)]|\nonumber\\
=&K_i\rho K_i^{\dagger},
\end{align}
which completes the proof. $\hfill\blacksquare$

In order to show the strong monotonicity of SIO, we need to prove
\begin{align}
C_m(\rho)\geq \sum_ip_i C_m(\frac{K_i\rho K_i^{\dagger}}{p_i}),
\end{align}
where $\{K_i\}$ is Kraus operators of arbitrary given SIO $\varepsilon_K$ and $p_i=\mathrm{Tr}(K_i\rho K_i^{\dagger})$.

Let $|\psi\rangle\in R(\rho)$ corresponding to the state $\rho$.  According to Lemma 1, there exists an SIO $\varepsilon_T=\{T_i\}$  such that
\begin{align}
\mathrm{Tr}(T_i|\psi\rangle\langle\psi| T_i^{\dagger})=\mathrm{Tr}(K_i\rho K_i^{\dagger})=p_i,
\end{align}
and $\frac{T_i|\psi\rangle\langle\psi| T_i^{\dagger}}{p_i}$ can be converted into $\frac{K_i\rho K_i^{\dagger}}{p_i}$ by SIO. Thus
\begin{align}
C_m(\rho)=&F(|\psi\rangle)\geq \sum_ip_iF(\frac{T_i|\psi\rangle\langle\psi| T_i^{\dagger}}{p_i})\nonumber\\
=&\sum_ip_iC_m(\frac{T_i|\psi\rangle\langle\psi| T_i^{\dagger}}{p_i})\geq \sum_ip_iC_m(\frac{K_i\rho K_i^{\dagger}}{p_i}),
\end{align}
where the first inequality results from the strong monotonicity of $C_m$, the second inequality is due to $\frac{T_i|\psi\rangle\langle\psi| T_i^{\dagger}}{p_i}\to\frac{K_i\rho K_i^{\dagger}}{p_i}$ by SIO and the monotonicity of $C_m$.

\bibliography{reference}

\begin{thebibliography}{80}%
\makeatletter
\providecommand \@ifxundefined [1]{%
 \@ifx{#1\undefined}
}%
\providecommand \@ifnum [1]{%
 \ifnum #1\expandafter \@firstoftwo
 \else \expandafter \@secondoftwo
 \fi
}%
\providecommand \@ifx [1]{%
 \ifx #1\expandafter \@firstoftwo
 \else \expandafter \@secondoftwo
 \fi
}%
\providecommand \natexlab [1]{#1}%
\providecommand \enquote  [1]{``#1''}%
\providecommand \bibnamefont  [1]{#1}%
\providecommand \bibfnamefont [1]{#1}%
\providecommand \citenamefont [1]{#1}%
\providecommand \href@noop [0]{\@secondoftwo}%
\providecommand \href [0]{\begingroup \@sanitize@url \@href}%
\providecommand \@href[1]{\@@startlink{#1}\@@href}%
\providecommand \@@href[1]{\endgroup#1\@@endlink}%
\providecommand \@sanitize@url [0]{\catcode `\\12\catcode `\$12\catcode
  `\&12\catcode `\#12\catcode `\^12\catcode `\_12\catcode `\%12\relax}%
\providecommand \@@startlink[1]{}%
\providecommand \@@endlink[0]{}%
\providecommand \url  [0]{\begingroup\@sanitize@url \@url }%
\providecommand \@url [1]{\endgroup\@href {#1}{\urlprefix }}%
\providecommand \urlprefix  [0]{URL }%
\providecommand \Eprint [0]{\href }%
\providecommand \doibase [0]{http://dx.doi.org/}%
\providecommand \selectlanguage [0]{\@gobble}%
\providecommand \bibinfo  [0]{\@secondoftwo}%
\providecommand \bibfield  [0]{\@secondoftwo}%
\providecommand \translation [1]{[#1]}%
\providecommand \BibitemOpen [0]{}%
\providecommand \bibitemStop [0]{}%
\providecommand \bibitemNoStop [0]{.\EOS\space}%
\providecommand \EOS [0]{\spacefactor3000\relax}%
\providecommand \BibitemShut  [1]{\csname bibitem#1\endcsname}%
\let\auto@bib@innerbib\@empty
\bibitem [{\citenamefont {Horodecki}\ \emph {et~al.}(2009)\citenamefont
  {Horodecki}, \citenamefont {Horodecki}, \citenamefont {Horodecki},\ and\
  \citenamefont {Horodecki}}]{entanglement}%
  \BibitemOpen
  \bibfield  {author} {\bibinfo {author} {\bibfnamefont {R.}~\bibnamefont
  {Horodecki}}, \bibinfo {author} {\bibfnamefont {P.}~\bibnamefont
  {Horodecki}}, \bibinfo {author} {\bibfnamefont {M.}~\bibnamefont
  {Horodecki}}, \ and\ \bibinfo {author} {\bibfnamefont {K.}~\bibnamefont
  {Horodecki}},\ }\href {\doibase 10.1103/RevModPhys.81.865} {\bibfield
  {journal} {\bibinfo  {journal} {Rev. Mod. Phys.}\ }\textbf {\bibinfo {volume}
  {81}},\ \bibinfo {pages} {865} (\bibinfo {year} {2009})}\BibitemShut
  {NoStop}%
\bibitem [{\citenamefont {Vedral}\ \emph {et~al.}(1997)\citenamefont {Vedral},
  \citenamefont {Plenio}, \citenamefont {Rippin},\ and\ \citenamefont
  {Knight}}]{entangle}%
  \BibitemOpen
  \bibfield  {author} {\bibinfo {author} {\bibfnamefont {V.}~\bibnamefont
  {Vedral}}, \bibinfo {author} {\bibfnamefont {M.~B.}\ \bibnamefont {Plenio}},
  \bibinfo {author} {\bibfnamefont {M.~A.}\ \bibnamefont {Rippin}}, \ and\
  \bibinfo {author} {\bibfnamefont {P.~L.}\ \bibnamefont {Knight}},\ }\href
  {\doibase 10.1103/PhysRevLett.78.2275} {\bibfield  {journal} {\bibinfo
  {journal} {Phys. Rev. Lett.}\ }\textbf {\bibinfo {volume} {78}},\ \bibinfo
  {pages} {2275} (\bibinfo {year} {1997})}\BibitemShut {NoStop}%
\bibitem [{\citenamefont {Girolami}\ and\ \citenamefont {Yadin}(2017)}]{e1}%
  \BibitemOpen
  \bibfield  {author} {\bibinfo {author} {\bibfnamefont {D.}~\bibnamefont
  {Girolami}}\ and\ \bibinfo {author} {\bibfnamefont {B.}~\bibnamefont
  {Yadin}},\ }\href
  {http://search.ebscohost.com/login.aspx?direct=true&db=a9h&AN=122552326&lang=zh-cn&site=eds-live}
  {\bibfield  {journal} {\bibinfo  {journal} {Entropy}\ }\textbf {\bibinfo
  {volume} {19}},\ \bibinfo {pages} {124} (\bibinfo {year} {2017})}\BibitemShut
  {NoStop}%
\bibitem [{\citenamefont {Bu}\ \emph {et~al.}(2016)\citenamefont {Bu},
  \citenamefont {Kumar},\ and\ \citenamefont {Wu}}]{e2}%
  \BibitemOpen
  \bibfield  {author} {\bibinfo {author} {\bibfnamefont {K.}~\bibnamefont
  {Bu}}, \bibinfo {author} {\bibfnamefont {A.}~\bibnamefont {Kumar}}, \ and\
  \bibinfo {author} {\bibfnamefont {J.}~\bibnamefont {Wu}},\ }\href@noop {}
  {\enquote {\bibinfo {title} {Bell-type inequality in quantum coherence theory
  as an entanglement witness},}\ } (\bibinfo {year} {2016}),\ \Eprint
  {http://arxiv.org/abs/1603.06322} {arXiv:1603.06322} \BibitemShut {NoStop}%
\bibitem [{\citenamefont {Chitambar}\ and\ \citenamefont {Hsieh}(2016)}]{e3}%
  \BibitemOpen
  \bibfield  {author} {\bibinfo {author} {\bibfnamefont {E.}~\bibnamefont
  {Chitambar}}\ and\ \bibinfo {author} {\bibfnamefont {M.-H.}\ \bibnamefont
  {Hsieh}},\ }\href {\doibase 10.1103/PhysRevLett.117.020402} {\bibfield
  {journal} {\bibinfo  {journal} {Phys. Rev. Lett.}\ }\textbf {\bibinfo
  {volume} {117}},\ \bibinfo {pages} {020402} (\bibinfo {year}
  {2016})}\BibitemShut {NoStop}%
\bibitem [{\citenamefont {Streltsov}\ \emph {et~al.}(2015)\citenamefont
  {Streltsov}, \citenamefont {Singh}, \citenamefont {Dhar}, \citenamefont
  {Bera},\ and\ \citenamefont {Adesso}}]{op2}%
  \BibitemOpen
  \bibfield  {author} {\bibinfo {author} {\bibfnamefont {A.}~\bibnamefont
  {Streltsov}}, \bibinfo {author} {\bibfnamefont {U.}~\bibnamefont {Singh}},
  \bibinfo {author} {\bibfnamefont {H.~S.}\ \bibnamefont {Dhar}}, \bibinfo
  {author} {\bibfnamefont {M.~N.}\ \bibnamefont {Bera}}, \ and\ \bibinfo
  {author} {\bibfnamefont {G.}~\bibnamefont {Adesso}},\ }\href {\doibase
  10.1103/PhysRevLett.115.020403} {\bibfield  {journal} {\bibinfo  {journal}
  {Phys. Rev. Lett.}\ }\textbf {\bibinfo {volume} {115}},\ \bibinfo {pages}
  {020403} (\bibinfo {year} {2015})}\BibitemShut {NoStop}%
\bibitem [{\citenamefont {Li}\ \emph {et~al.}(2005)\citenamefont {Li},
  \citenamefont {Xiong},\ and\ \citenamefont {Zubairy}}]{e4}%
  \BibitemOpen
  \bibfield  {author} {\bibinfo {author} {\bibfnamefont {F.-l.}\ \bibnamefont
  {Li}}, \bibinfo {author} {\bibfnamefont {H.}~\bibnamefont {Xiong}}, \ and\
  \bibinfo {author} {\bibfnamefont {M.~S.}\ \bibnamefont {Zubairy}},\ }\href
  {\doibase 10.1103/PhysRevA.72.010303} {\bibfield  {journal} {\bibinfo
  {journal} {Phys. Rev. A}\ }\textbf {\bibinfo {volume} {72}},\ \bibinfo
  {pages} {010303} (\bibinfo {year} {2005})}\BibitemShut {NoStop}%
\bibitem [{\citenamefont {Ollivier}\ and\ \citenamefont
  {Zurek}(2001)}]{discord}%
  \BibitemOpen
  \bibfield  {author} {\bibinfo {author} {\bibfnamefont {H.}~\bibnamefont
  {Ollivier}}\ and\ \bibinfo {author} {\bibfnamefont {W.~H.}\ \bibnamefont
  {Zurek}},\ }\href {\doibase 10.1103/PhysRevLett.88.017901} {\bibfield
  {journal} {\bibinfo  {journal} {Phys. Rev. Lett.}\ }\textbf {\bibinfo
  {volume} {88}},\ \bibinfo {pages} {017901} (\bibinfo {year}
  {2001})}\BibitemShut {NoStop}%
\bibitem [{\citenamefont {Ma}\ \emph {et~al.}(2016)\citenamefont {Ma},
  \citenamefont {Yadin}, \citenamefont {Girolami}, \citenamefont {Vedral},\
  and\ \citenamefont {Gu}}]{cor}%
  \BibitemOpen
  \bibfield  {author} {\bibinfo {author} {\bibfnamefont {J.}~\bibnamefont
  {Ma}}, \bibinfo {author} {\bibfnamefont {B.}~\bibnamefont {Yadin}}, \bibinfo
  {author} {\bibfnamefont {D.}~\bibnamefont {Girolami}}, \bibinfo {author}
  {\bibfnamefont {V.}~\bibnamefont {Vedral}}, \ and\ \bibinfo {author}
  {\bibfnamefont {M.}~\bibnamefont {Gu}},\ }\href {\doibase
  10.1103/PhysRevLett.116.160407} {\bibfield  {journal} {\bibinfo  {journal}
  {Phys. Rev. Lett.}\ }\textbf {\bibinfo {volume} {116}},\ \bibinfo {pages}
  {160407} (\bibinfo {year} {2016})}\BibitemShut {NoStop}%
\bibitem [{\citenamefont {Arthurs}\ and\ \citenamefont
  {Goodman}(1988)}]{correlation}%
  \BibitemOpen
  \bibfield  {author} {\bibinfo {author} {\bibfnamefont {E.}~\bibnamefont
  {Arthurs}}\ and\ \bibinfo {author} {\bibfnamefont {M.~S.}\ \bibnamefont
  {Goodman}},\ }\href {\doibase 10.1103/PhysRevLett.60.2447} {\bibfield
  {journal} {\bibinfo  {journal} {Phys. Rev. Lett.}\ }\textbf {\bibinfo
  {volume} {60}},\ \bibinfo {pages} {2447} (\bibinfo {year}
  {1988})}\BibitemShut {NoStop}%
\bibitem [{\citenamefont {Sun}\ \emph {et~al.}(2017)\citenamefont {Sun},
  \citenamefont {Mao},\ and\ \citenamefont {Luo}}]{Sun_2017}%
  \BibitemOpen
  \bibfield  {author} {\bibinfo {author} {\bibfnamefont {Y.}~\bibnamefont
  {Sun}}, \bibinfo {author} {\bibfnamefont {Y.}~\bibnamefont {Mao}}, \ and\
  \bibinfo {author} {\bibfnamefont {S.}~\bibnamefont {Luo}},\ }\href {\doibase
  10.1209/0295-5075/118/60007} {\bibfield  {journal} {\bibinfo  {journal}
  {Europhys. Lett.}\ }\textbf {\bibinfo {volume} {118}},\ \bibinfo {pages}
  {60007} (\bibinfo {year} {2017})}\BibitemShut {NoStop}%
\bibitem [{\citenamefont {Mondal}\ \emph {et~al.}(2017)\citenamefont {Mondal},
  \citenamefont {Pramanik},\ and\ \citenamefont {Pati}}]{non1}%
  \BibitemOpen
  \bibfield  {author} {\bibinfo {author} {\bibfnamefont {D.}~\bibnamefont
  {Mondal}}, \bibinfo {author} {\bibfnamefont {T.}~\bibnamefont {Pramanik}}, \
  and\ \bibinfo {author} {\bibfnamefont {A.~K.}\ \bibnamefont {Pati}},\ }\href
  {\doibase 10.1103/PhysRevA.95.010301} {\bibfield  {journal} {\bibinfo
  {journal} {Phys. Rev. A}\ }\textbf {\bibinfo {volume} {95}},\ \bibinfo
  {pages} {010301} (\bibinfo {year} {2017})}\BibitemShut {NoStop}%
\bibitem [{\citenamefont {Mondal}\ and\ \citenamefont
  {Mukhopadhyay}(2015)}]{non2}%
  \BibitemOpen
  \bibfield  {author} {\bibinfo {author} {\bibfnamefont {D.}~\bibnamefont
  {Mondal}}\ and\ \bibinfo {author} {\bibfnamefont {C.}~\bibnamefont
  {Mukhopadhyay}},\ }\href@noop {} {\enquote {\bibinfo {title} {Steerability of
  quantum coherence in accelerated frame},}\ } (\bibinfo {year} {2015}),\
  \Eprint {http://arxiv.org/abs/1510.07556} {arXiv:1510.07556} \BibitemShut
  {NoStop}%
\bibitem [{\citenamefont {Hu}\ \emph {et~al.}(2018)\citenamefont {Hu},
  \citenamefont {Wang},\ and\ \citenamefont {Fan}}]{non3}%
  \BibitemOpen
  \bibfield  {author} {\bibinfo {author} {\bibfnamefont {M.-L.}\ \bibnamefont
  {Hu}}, \bibinfo {author} {\bibfnamefont {X.-M.}\ \bibnamefont {Wang}}, \ and\
  \bibinfo {author} {\bibfnamefont {H.}~\bibnamefont {Fan}},\ }\href {\doibase
  10.1103/PhysRevA.98.032317} {\bibfield  {journal} {\bibinfo  {journal} {Phys.
  Rev. A}\ }\textbf {\bibinfo {volume} {98}},\ \bibinfo {pages} {032317}
  (\bibinfo {year} {2018})}\BibitemShut {NoStop}%
\bibitem [{\citenamefont {Du}\ \emph {et~al.}(2017)\citenamefont {Du},
  \citenamefont {Wang},\ and\ \citenamefont {Ye}}]{non4}%
  \BibitemOpen
  \bibfield  {author} {\bibinfo {author} {\bibfnamefont {M.-M.}\ \bibnamefont
  {Du}}, \bibinfo {author} {\bibfnamefont {D.}~\bibnamefont {Wang}}, \ and\
  \bibinfo {author} {\bibfnamefont {L.}~\bibnamefont {Ye}},\ }\href {\doibase
  10.1007/s11128-017-1663-2} {\bibfield  {journal} {\bibinfo  {journal}
  {Quantum Inf. Process.}\ }\textbf {\bibinfo {volume} {16}},\ \bibinfo {pages}
  {218} (\bibinfo {year} {2017})}\BibitemShut {NoStop}%
\bibitem [{\citenamefont {Datta}\ and\ \citenamefont {Majumdar}(2018)}]{non5}%
  \BibitemOpen
  \bibfield  {author} {\bibinfo {author} {\bibfnamefont {S.}~\bibnamefont
  {Datta}}\ and\ \bibinfo {author} {\bibfnamefont {A.~S.}\ \bibnamefont
  {Majumdar}},\ }\href {\doibase 10.1103/PhysRevA.98.042311} {\bibfield
  {journal} {\bibinfo  {journal} {Phys. Rev. A}\ }\textbf {\bibinfo {volume}
  {98}},\ \bibinfo {pages} {042311} (\bibinfo {year} {2018})}\BibitemShut
  {NoStop}%
\bibitem [{\citenamefont {Marvian}\ and\ \citenamefont {Spekkens}(2014)}]{as}%
  \BibitemOpen
  \bibfield  {author} {\bibinfo {author} {\bibfnamefont {I.}~\bibnamefont
  {Marvian}}\ and\ \bibinfo {author} {\bibfnamefont {R.}~\bibnamefont
  {Spekkens}},\ }\href {\doibase 10.1038/ncomms4821} {\bibfield  {journal}
  {\bibinfo  {journal} {Nat. Commun.}\ }\textbf {\bibinfo {volume} {5}},\
  \bibinfo {pages} {3821} (\bibinfo {year} {2014})}\BibitemShut {NoStop}%
\bibitem [{\citenamefont {Marvian}\ \emph {et~al.}(2016)\citenamefont
  {Marvian}, \citenamefont {Spekkens},\ and\ \citenamefont {Zanardi}}]{as1}%
  \BibitemOpen
  \bibfield  {author} {\bibinfo {author} {\bibfnamefont {I.}~\bibnamefont
  {Marvian}}, \bibinfo {author} {\bibfnamefont {R.~W.}\ \bibnamefont
  {Spekkens}}, \ and\ \bibinfo {author} {\bibfnamefont {P.}~\bibnamefont
  {Zanardi}},\ }\href {\doibase 10.1103/PhysRevA.93.052331} {\bibfield
  {journal} {\bibinfo  {journal} {Phys. Rev. A}\ }\textbf {\bibinfo {volume}
  {93}},\ \bibinfo {pages} {052331} (\bibinfo {year} {2016})}\BibitemShut
  {NoStop}%
\bibitem [{\citenamefont {Piani}\ \emph {et~al.}(2016)\citenamefont {Piani},
  \citenamefont {Cianciaruso}, \citenamefont {Bromley}, \citenamefont {Napoli},
  \citenamefont {Johnston},\ and\ \citenamefont {Adesso}}]{as2}%
  \BibitemOpen
  \bibfield  {author} {\bibinfo {author} {\bibfnamefont {M.}~\bibnamefont
  {Piani}}, \bibinfo {author} {\bibfnamefont {M.}~\bibnamefont {Cianciaruso}},
  \bibinfo {author} {\bibfnamefont {T.~R.}\ \bibnamefont {Bromley}}, \bibinfo
  {author} {\bibfnamefont {C.}~\bibnamefont {Napoli}}, \bibinfo {author}
  {\bibfnamefont {N.}~\bibnamefont {Johnston}}, \ and\ \bibinfo {author}
  {\bibfnamefont {G.}~\bibnamefont {Adesso}},\ }\href {\doibase
  10.1103/PhysRevA.93.042107} {\bibfield  {journal} {\bibinfo  {journal} {Phys.
  Rev. A}\ }\textbf {\bibinfo {volume} {93}},\ \bibinfo {pages} {042107}
  (\bibinfo {year} {2016})}\BibitemShut {NoStop}%
\bibitem [{\citenamefont {Lostaglio}\ \emph {et~al.}(2015)\citenamefont
  {Lostaglio}, \citenamefont {Korzekwa}, \citenamefont {Jennings},\ and\
  \citenamefont {Rudolph}}]{ther1}%
  \BibitemOpen
  \bibfield  {author} {\bibinfo {author} {\bibfnamefont {M.}~\bibnamefont
  {Lostaglio}}, \bibinfo {author} {\bibfnamefont {K.}~\bibnamefont {Korzekwa}},
  \bibinfo {author} {\bibfnamefont {D.}~\bibnamefont {Jennings}}, \ and\
  \bibinfo {author} {\bibfnamefont {T.}~\bibnamefont {Rudolph}},\ }\href
  {\doibase 10.1103/PhysRevX.5.021001} {\bibfield  {journal} {\bibinfo
  {journal} {Phys. Rev. X}\ }\textbf {\bibinfo {volume} {5}},\ \bibinfo {pages}
  {021001} (\bibinfo {year} {2015})}\BibitemShut {NoStop}%
\bibitem [{\citenamefont {Gour}\ \emph {et~al.}(2015)\citenamefont {Gour},
  \citenamefont {M{\"u}ller}, \citenamefont {Narasimhachar}, \citenamefont
  {Spekkens},\ and\ \citenamefont {Halpern}}]{ther2}%
  \BibitemOpen
  \bibfield  {author} {\bibinfo {author} {\bibfnamefont {G.}~\bibnamefont
  {Gour}}, \bibinfo {author} {\bibfnamefont {M.~P.}\ \bibnamefont
  {M{\"u}ller}}, \bibinfo {author} {\bibfnamefont {V.}~\bibnamefont
  {Narasimhachar}}, \bibinfo {author} {\bibfnamefont {R.~W.}\ \bibnamefont
  {Spekkens}}, \ and\ \bibinfo {author} {\bibfnamefont {N.~Y.}\ \bibnamefont
  {Halpern}},\ }\href {\doibase https://doi.org/10.1016/j.physrep.2015.04.003}
  {\bibfield  {journal} {\bibinfo  {journal} {Phys. Rep.}\ }\textbf {\bibinfo
  {volume} {583}},\ \bibinfo {pages} {1 } (\bibinfo {year} {2015})}\BibitemShut
  {NoStop}%
\bibitem [{\citenamefont {Rybak}\ \emph {et~al.}(2011)\citenamefont {Rybak},
  \citenamefont {Amaran}, \citenamefont {Levin}, \citenamefont {Tomza},
  \citenamefont {Moszynski}, \citenamefont {Kosloff}, \citenamefont {Koch},\
  and\ \citenamefont {Amitay}}]{ther3}%
  \BibitemOpen
  \bibfield  {author} {\bibinfo {author} {\bibfnamefont {L.}~\bibnamefont
  {Rybak}}, \bibinfo {author} {\bibfnamefont {S.}~\bibnamefont {Amaran}},
  \bibinfo {author} {\bibfnamefont {L.}~\bibnamefont {Levin}}, \bibinfo
  {author} {\bibfnamefont {M.}~\bibnamefont {Tomza}}, \bibinfo {author}
  {\bibfnamefont {R.}~\bibnamefont {Moszynski}}, \bibinfo {author}
  {\bibfnamefont {R.}~\bibnamefont {Kosloff}}, \bibinfo {author} {\bibfnamefont
  {C.~P.}\ \bibnamefont {Koch}}, \ and\ \bibinfo {author} {\bibfnamefont
  {Z.}~\bibnamefont {Amitay}},\ }\href {\doibase
  10.1103/PhysRevLett.107.273001} {\bibfield  {journal} {\bibinfo  {journal}
  {Phys. Rev. Lett.}\ }\textbf {\bibinfo {volume} {107}},\ \bibinfo {pages}
  {273001} (\bibinfo {year} {2011})}\BibitemShut {NoStop}%
\bibitem [{\citenamefont {Misra}\ \emph {et~al.}(2016)\citenamefont {Misra},
  \citenamefont {Singh}, \citenamefont {Bhattacharya},\ and\ \citenamefont
  {Pati}}]{ther4}%
  \BibitemOpen
  \bibfield  {author} {\bibinfo {author} {\bibfnamefont {A.}~\bibnamefont
  {Misra}}, \bibinfo {author} {\bibfnamefont {U.}~\bibnamefont {Singh}},
  \bibinfo {author} {\bibfnamefont {S.}~\bibnamefont {Bhattacharya}}, \ and\
  \bibinfo {author} {\bibfnamefont {A.~K.}\ \bibnamefont {Pati}},\ }\href
  {\doibase 10.1103/PhysRevA.93.052335} {\bibfield  {journal} {\bibinfo
  {journal} {Phys. Rev. A}\ }\textbf {\bibinfo {volume} {93}},\ \bibinfo
  {pages} {052335} (\bibinfo {year} {2016})}\BibitemShut {NoStop}%
\bibitem [{\citenamefont {Scully}\ \emph {et~al.}(2011)\citenamefont {Scully},
  \citenamefont {Chapin}, \citenamefont {Dorfman}, \citenamefont {Kim},\ and\
  \citenamefont {Svidzinsky}}]{ther5}%
  \BibitemOpen
  \bibfield  {author} {\bibinfo {author} {\bibfnamefont {M.~O.}\ \bibnamefont
  {Scully}}, \bibinfo {author} {\bibfnamefont {K.~R.}\ \bibnamefont {Chapin}},
  \bibinfo {author} {\bibfnamefont {K.~E.}\ \bibnamefont {Dorfman}}, \bibinfo
  {author} {\bibfnamefont {M.~B.}\ \bibnamefont {Kim}}, \ and\ \bibinfo
  {author} {\bibfnamefont {A.}~\bibnamefont {Svidzinsky}},\ }\href {\doibase
  10.1073/pnas.1110234108} {\bibfield  {journal} {\bibinfo  {journal} {Proc.
  Natl. Acad. Sci. U. S. A.}\ }\textbf {\bibinfo {volume} {108}},\ \bibinfo
  {pages} {15097} (\bibinfo {year} {2011})}\BibitemShut {NoStop}%
\bibitem [{\citenamefont {Brandner}\ \emph {et~al.}(2017)\citenamefont
  {Brandner}, \citenamefont {Bauer},\ and\ \citenamefont {Seifert}}]{ther6}%
  \BibitemOpen
  \bibfield  {author} {\bibinfo {author} {\bibfnamefont {K.}~\bibnamefont
  {Brandner}}, \bibinfo {author} {\bibfnamefont {M.}~\bibnamefont {Bauer}}, \
  and\ \bibinfo {author} {\bibfnamefont {U.}~\bibnamefont {Seifert}},\ }\href
  {\doibase 10.1103/PhysRevLett.119.170602} {\bibfield  {journal} {\bibinfo
  {journal} {Phys. Rev. Lett.}\ }\textbf {\bibinfo {volume} {119}},\ \bibinfo
  {pages} {170602} (\bibinfo {year} {2017})}\BibitemShut {NoStop}%
\bibitem [{\citenamefont {Lloyd}(2011)}]{qb1}%
  \BibitemOpen
  \bibfield  {author} {\bibinfo {author} {\bibfnamefont {S.}~\bibnamefont
  {Lloyd}},\ }\href {\doibase 10.1088/1742-6596/302/1/012037} {\bibfield
  {journal} {\bibinfo  {journal} {J. Phys.: Conf. Ser.}\ }\textbf {\bibinfo
  {volume} {302}},\ \bibinfo {pages} {012037} (\bibinfo {year}
  {2011})}\BibitemShut {NoStop}%
\bibitem [{\citenamefont {Engel}\ \emph {et~al.}(2007)\citenamefont {Engel},
  \citenamefont {Calhoun}, \citenamefont {Read}, \citenamefont {Ahn},
  \citenamefont {Mancal}, \citenamefont {Cheng}, \citenamefont {Blankenship},\
  and\ \citenamefont {Fleming}}]{qb}%
  \BibitemOpen
  \bibfield  {author} {\bibinfo {author} {\bibfnamefont {G.}~\bibnamefont
  {Engel}}, \bibinfo {author} {\bibfnamefont {T.}~\bibnamefont {Calhoun}},
  \bibinfo {author} {\bibfnamefont {E.}~\bibnamefont {Read}}, \bibinfo {author}
  {\bibfnamefont {T.-K.}\ \bibnamefont {Ahn}}, \bibinfo {author} {\bibfnamefont
  {T.}~\bibnamefont {Mancal}}, \bibinfo {author} {\bibfnamefont {Y.-C.}\
  \bibnamefont {Cheng}}, \bibinfo {author} {\bibfnamefont {R.}~\bibnamefont
  {Blankenship}}, \ and\ \bibinfo {author} {\bibfnamefont {G.}~\bibnamefont
  {Fleming}},\ }\href {\doibase 10.1038/nature05678} {\bibfield  {journal}
  {\bibinfo  {journal} {Nature(London)}\ }\textbf {\bibinfo {volume} {446}},\
  \bibinfo {pages} {782} (\bibinfo {year} {2007})}\BibitemShut {NoStop}%
\bibitem [{\citenamefont {Gauger}\ \emph {et~al.}(2011)\citenamefont {Gauger},
  \citenamefont {Rieper}, \citenamefont {Morton}, \citenamefont {Benjamin},\
  and\ \citenamefont {Vedral}}]{qb2}%
  \BibitemOpen
  \bibfield  {author} {\bibinfo {author} {\bibfnamefont {E.~M.}\ \bibnamefont
  {Gauger}}, \bibinfo {author} {\bibfnamefont {E.}~\bibnamefont {Rieper}},
  \bibinfo {author} {\bibfnamefont {J.~J.~L.}\ \bibnamefont {Morton}}, \bibinfo
  {author} {\bibfnamefont {S.~C.}\ \bibnamefont {Benjamin}}, \ and\ \bibinfo
  {author} {\bibfnamefont {V.}~\bibnamefont {Vedral}},\ }\href {\doibase
  10.1103/PhysRevLett.106.040503} {\bibfield  {journal} {\bibinfo  {journal}
  {Phys. Rev. Lett.}\ }\textbf {\bibinfo {volume} {106}},\ \bibinfo {pages}
  {040503} (\bibinfo {year} {2011})}\BibitemShut {NoStop}%
\bibitem [{\citenamefont {Li}\ \emph {et~al.}(2012)\citenamefont {Li},
  \citenamefont {Lambert}, \citenamefont {Chen}, \citenamefont {Chen},\ and\
  \citenamefont {Nori}}]{qb3}%
  \BibitemOpen
  \bibfield  {author} {\bibinfo {author} {\bibfnamefont {C.-M.}\ \bibnamefont
  {Li}}, \bibinfo {author} {\bibfnamefont {N.}~\bibnamefont {Lambert}},
  \bibinfo {author} {\bibfnamefont {Y.-N.}\ \bibnamefont {Chen}}, \bibinfo
  {author} {\bibfnamefont {G.-Y.}\ \bibnamefont {Chen}}, \ and\ \bibinfo
  {author} {\bibfnamefont {F.}~\bibnamefont {Nori}},\ }\href {\doibase
  10.1038/srep00885} {\bibfield  {journal} {\bibinfo  {journal} {Sci. Rep.}\
  }\textbf {\bibinfo {volume} {2}},\ \bibinfo {pages} {885} (\bibinfo {year}
  {2012})}\BibitemShut {NoStop}%
\bibitem [{\citenamefont {Wang}\ \emph {et~al.}(2018)\citenamefont {Wang},
  \citenamefont {Wu}, \citenamefont {Cui},\ and\ \citenamefont {Wang}}]{qm}%
  \BibitemOpen
  \bibfield  {author} {\bibinfo {author} {\bibfnamefont {Z.}~\bibnamefont
  {Wang}}, \bibinfo {author} {\bibfnamefont {W.}~\bibnamefont {Wu}}, \bibinfo
  {author} {\bibfnamefont {G.}~\bibnamefont {Cui}}, \ and\ \bibinfo {author}
  {\bibfnamefont {J.}~\bibnamefont {Wang}},\ }\href {\doibase
  10.1088/1367-2630/aab03a} {\bibfield  {journal} {\bibinfo  {journal} {New J.
  Phys.}\ }\textbf {\bibinfo {volume} {20}},\ \bibinfo {pages} {033034}
  (\bibinfo {year} {2018})}\BibitemShut {NoStop}%
\bibitem [{\citenamefont {Braunstein}\ and\ \citenamefont {Caves}(1994)}]{m1}%
  \BibitemOpen
  \bibfield  {author} {\bibinfo {author} {\bibfnamefont {S.~L.}\ \bibnamefont
  {Braunstein}}\ and\ \bibinfo {author} {\bibfnamefont {C.~M.}\ \bibnamefont
  {Caves}},\ }\href {\doibase 10.1103/PhysRevLett.72.3439} {\bibfield
  {journal} {\bibinfo  {journal} {Phys. Rev. Lett.}\ }\textbf {\bibinfo
  {volume} {72}},\ \bibinfo {pages} {3439} (\bibinfo {year}
  {1994})}\BibitemShut {NoStop}%
\bibitem [{\citenamefont {Giovannetti}\ \emph {et~al.}(2006)\citenamefont
  {Giovannetti}, \citenamefont {Lloyd},\ and\ \citenamefont {Maccone}}]{m2}%
  \BibitemOpen
  \bibfield  {author} {\bibinfo {author} {\bibfnamefont {V.}~\bibnamefont
  {Giovannetti}}, \bibinfo {author} {\bibfnamefont {S.}~\bibnamefont {Lloyd}},
  \ and\ \bibinfo {author} {\bibfnamefont {L.}~\bibnamefont {Maccone}},\ }\href
  {\doibase 10.1103/PhysRevLett.96.010401} {\bibfield  {journal} {\bibinfo
  {journal} {Phys. Rev. Lett.}\ }\textbf {\bibinfo {volume} {96}},\ \bibinfo
  {pages} {010401} (\bibinfo {year} {2006})}\BibitemShut {NoStop}%
\bibitem [{\citenamefont {Zhang}\ \emph
  {et~al.}(2017{\natexlab{a}})\citenamefont {Zhang}, \citenamefont {Yadin},
  \citenamefont {Hou}, \citenamefont {Cao}, \citenamefont {Liu}, \citenamefont
  {Huang}, \citenamefont {Maity}, \citenamefont {Vedral}, \citenamefont {Li},
  \citenamefont {Guo},\ and\ \citenamefont {Girolami}}]{m3}%
  \BibitemOpen
  \bibfield  {author} {\bibinfo {author} {\bibfnamefont {C.}~\bibnamefont
  {Zhang}}, \bibinfo {author} {\bibfnamefont {B.}~\bibnamefont {Yadin}},
  \bibinfo {author} {\bibfnamefont {Z.-B.}\ \bibnamefont {Hou}}, \bibinfo
  {author} {\bibfnamefont {H.}~\bibnamefont {Cao}}, \bibinfo {author}
  {\bibfnamefont {B.-H.}\ \bibnamefont {Liu}}, \bibinfo {author} {\bibfnamefont
  {Y.-F.}\ \bibnamefont {Huang}}, \bibinfo {author} {\bibfnamefont
  {R.}~\bibnamefont {Maity}}, \bibinfo {author} {\bibfnamefont
  {V.}~\bibnamefont {Vedral}}, \bibinfo {author} {\bibfnamefont {C.-F.}\
  \bibnamefont {Li}}, \bibinfo {author} {\bibfnamefont {G.-C.}\ \bibnamefont
  {Guo}}, \ and\ \bibinfo {author} {\bibfnamefont {D.}~\bibnamefont
  {Girolami}},\ }\href {\doibase 10.1103/PhysRevA.96.042327} {\bibfield
  {journal} {\bibinfo  {journal} {Phys. Rev. A}\ }\textbf {\bibinfo {volume}
  {96}},\ \bibinfo {pages} {042327} (\bibinfo {year}
  {2017}{\natexlab{a}})}\BibitemShut {NoStop}%
\bibitem [{\citenamefont {Giorda}\ and\ \citenamefont {Allegra}(2017)}]{m4}%
  \BibitemOpen
  \bibfield  {author} {\bibinfo {author} {\bibfnamefont {P.}~\bibnamefont
  {Giorda}}\ and\ \bibinfo {author} {\bibfnamefont {M.}~\bibnamefont
  {Allegra}},\ }\href {\doibase 10.1088/1751-8121/aa9808} {\bibfield  {journal}
  {\bibinfo  {journal} {J. Phys. A: Math. Theor.}\ }\textbf {\bibinfo {volume}
  {51}},\ \bibinfo {pages} {025302} (\bibinfo {year} {2017})}\BibitemShut
  {NoStop}%
\bibitem [{\citenamefont {Braun}\ \emph {et~al.}(2015)\citenamefont {Braun},
  \citenamefont {Friesdorf}, \citenamefont {Hodgman}, \citenamefont
  {Schreiber}, \citenamefont {Ronzheimer}, \citenamefont {Riera}, \citenamefont
  {del Rey}, \citenamefont {Bloch}, \citenamefont {Eisert},\ and\ \citenamefont
  {Schneider}}]{trans1}%
  \BibitemOpen
  \bibfield  {author} {\bibinfo {author} {\bibfnamefont {S.}~\bibnamefont
  {Braun}}, \bibinfo {author} {\bibfnamefont {M.}~\bibnamefont {Friesdorf}},
  \bibinfo {author} {\bibfnamefont {S.~S.}\ \bibnamefont {Hodgman}}, \bibinfo
  {author} {\bibfnamefont {M.}~\bibnamefont {Schreiber}}, \bibinfo {author}
  {\bibfnamefont {J.~P.}\ \bibnamefont {Ronzheimer}}, \bibinfo {author}
  {\bibfnamefont {A.}~\bibnamefont {Riera}}, \bibinfo {author} {\bibfnamefont
  {M.}~\bibnamefont {del Rey}}, \bibinfo {author} {\bibfnamefont
  {I.}~\bibnamefont {Bloch}}, \bibinfo {author} {\bibfnamefont
  {J.}~\bibnamefont {Eisert}}, \ and\ \bibinfo {author} {\bibfnamefont
  {U.}~\bibnamefont {Schneider}},\ }\href {\doibase 10.1073/pnas.1408861112}
  {\bibfield  {journal} {\bibinfo  {journal} {Proc. Natl. Acad. Sci. U. S. A.}\
  }\textbf {\bibinfo {volume} {112}},\ \bibinfo {pages} {3641} (\bibinfo {year}
  {2015})}\BibitemShut {NoStop}%
\bibitem [{\citenamefont {Malvezzi}\ \emph {et~al.}(2016)\citenamefont
  {Malvezzi}, \citenamefont {Karpat}, \citenamefont {\ifmmode~\mbox{\c{C}}\else
  \c{C}\fi{}akmak}, \citenamefont {Fanchini}, \citenamefont {Debarba},\ and\
  \citenamefont {Vianna}}]{trans2}%
  \BibitemOpen
  \bibfield  {author} {\bibinfo {author} {\bibfnamefont {A.~L.}\ \bibnamefont
  {Malvezzi}}, \bibinfo {author} {\bibfnamefont {G.}~\bibnamefont {Karpat}},
  \bibinfo {author} {\bibfnamefont {B.}~\bibnamefont
  {\ifmmode~\mbox{\c{C}}\else \c{C}\fi{}akmak}}, \bibinfo {author}
  {\bibfnamefont {F.~F.}\ \bibnamefont {Fanchini}}, \bibinfo {author}
  {\bibfnamefont {T.}~\bibnamefont {Debarba}}, \ and\ \bibinfo {author}
  {\bibfnamefont {R.~O.}\ \bibnamefont {Vianna}},\ }\href {\doibase
  10.1103/PhysRevB.93.184428} {\bibfield  {journal} {\bibinfo  {journal} {Phys.
  Rev. B}\ }\textbf {\bibinfo {volume} {93}},\ \bibinfo {pages} {184428}
  (\bibinfo {year} {2016})}\BibitemShut {NoStop}%
\bibitem [{\citenamefont {Li}\ and\ \citenamefont {Lin}(2016)}]{trans3}%
  \BibitemOpen
  \bibfield  {author} {\bibinfo {author} {\bibfnamefont {Y.-C.}\ \bibnamefont
  {Li}}\ and\ \bibinfo {author} {\bibfnamefont {H.-Q.}\ \bibnamefont {Lin}},\
  }\href {\doibase 10.1038/srep26365} {\bibfield  {journal} {\bibinfo
  {journal} {Sci. Rep.}\ }\textbf {\bibinfo {volume} {6}},\ \bibinfo {pages}
  {26365} (\bibinfo {year} {2016})}\BibitemShut {NoStop}%
\bibitem [{\citenamefont {Girolami}(2014{\natexlab{a}})}]{trans4}%
  \BibitemOpen
  \bibfield  {author} {\bibinfo {author} {\bibfnamefont {D.}~\bibnamefont
  {Girolami}},\ }\href {\doibase 10.1103/PhysRevLett.113.170401} {\bibfield
  {journal} {\bibinfo  {journal} {Phys. Rev. Lett.}\ }\textbf {\bibinfo
  {volume} {113}},\ \bibinfo {pages} {170401} (\bibinfo {year}
  {2014}{\natexlab{a}})}\BibitemShut {NoStop}%
\bibitem [{\citenamefont {Karpat}\ \emph {et~al.}(2014)\citenamefont {Karpat},
  \citenamefont {\ifmmode~\mbox{\c{C}}\else \c{C}\fi{}akmak},\ and\
  \citenamefont {Fanchini}}]{trans5}%
  \BibitemOpen
  \bibfield  {author} {\bibinfo {author} {\bibfnamefont {G.}~\bibnamefont
  {Karpat}}, \bibinfo {author} {\bibfnamefont {B.}~\bibnamefont
  {\ifmmode~\mbox{\c{C}}\else \c{C}\fi{}akmak}}, \ and\ \bibinfo {author}
  {\bibfnamefont {F.~F.}\ \bibnamefont {Fanchini}},\ }\href {\doibase
  10.1103/PhysRevB.90.104431} {\bibfield  {journal} {\bibinfo  {journal} {Phys.
  Rev. B}\ }\textbf {\bibinfo {volume} {90}},\ \bibinfo {pages} {104431}
  (\bibinfo {year} {2014})}\BibitemShut {NoStop}%
\bibitem [{\citenamefont {Streltsov}\ \emph {et~al.}(2017)\citenamefont
  {Streltsov}, \citenamefont {Adesso},\ and\ \citenamefont {Plenio}}]{qc2}%
  \BibitemOpen
  \bibfield  {author} {\bibinfo {author} {\bibfnamefont {A.}~\bibnamefont
  {Streltsov}}, \bibinfo {author} {\bibfnamefont {G.}~\bibnamefont {Adesso}}, \
  and\ \bibinfo {author} {\bibfnamefont {M.~B.}\ \bibnamefont {Plenio}},\
  }\href {\doibase 10.1103/RevModPhys.89.041003} {\bibfield  {journal}
  {\bibinfo  {journal} {Rev. Mod. Phys.}\ }\textbf {\bibinfo {volume} {89}},\
  \bibinfo {pages} {041003} (\bibinfo {year} {2017})}\BibitemShut {NoStop}%
\bibitem [{\citenamefont {Chitambar}\ and\ \citenamefont {Gour}(2019)}]{qrt1}%
  \BibitemOpen
  \bibfield  {author} {\bibinfo {author} {\bibfnamefont {E.}~\bibnamefont
  {Chitambar}}\ and\ \bibinfo {author} {\bibfnamefont {G.}~\bibnamefont
  {Gour}},\ }\href {\doibase 10.1103/RevModPhys.91.025001} {\bibfield
  {journal} {\bibinfo  {journal} {Rev. Mod. Phys.}\ }\textbf {\bibinfo {volume}
  {91}},\ \bibinfo {pages} {025001} (\bibinfo {year} {2019})}\BibitemShut
  {NoStop}%
\bibitem [{\citenamefont {Baumgratz}\ \emph {et~al.}(2014)\citenamefont
  {Baumgratz}, \citenamefont {Cramer},\ and\ \citenamefont {Plenio}}]{qc1}%
  \BibitemOpen
  \bibfield  {author} {\bibinfo {author} {\bibfnamefont {T.}~\bibnamefont
  {Baumgratz}}, \bibinfo {author} {\bibfnamefont {M.}~\bibnamefont {Cramer}}, \
  and\ \bibinfo {author} {\bibfnamefont {M.~B.}\ \bibnamefont {Plenio}},\
  }\href {\doibase 10.1103/PhysRevLett.113.140401} {\bibfield  {journal}
  {\bibinfo  {journal} {Phys. Rev. Lett.}\ }\textbf {\bibinfo {volume} {113}},\
  \bibinfo {pages} {140401} (\bibinfo {year} {2014})}\BibitemShut {NoStop}%
\bibitem [{\citenamefont {Aberg}()}]{roof1}%
  \BibitemOpen
  \bibfield  {author} {\bibinfo {author} {\bibfnamefont {J.}~\bibnamefont
  {Aberg}},\ }\href@noop {} {\ }\BibitemShut {NoStop}%
\bibitem [{\citenamefont {Chitambar}\ and\ \citenamefont
  {Gour}(2016{\natexlab{a}})}]{free}%
  \BibitemOpen
  \bibfield  {author} {\bibinfo {author} {\bibfnamefont {E.}~\bibnamefont
  {Chitambar}}\ and\ \bibinfo {author} {\bibfnamefont {G.}~\bibnamefont
  {Gour}},\ }\href {\doibase 10.1103/PhysRevA.94.052336} {\bibfield  {journal}
  {\bibinfo  {journal} {Phys. Rev. A}\ }\textbf {\bibinfo {volume} {94}},\
  \bibinfo {pages} {052336} (\bibinfo {year} {2016}{\natexlab{a}})}\BibitemShut
  {NoStop}%
\bibitem [{\citenamefont {Marvian}\ and\ \citenamefont
  {Spekkens}(2016)}]{speak}%
  \BibitemOpen
  \bibfield  {author} {\bibinfo {author} {\bibfnamefont {I.}~\bibnamefont
  {Marvian}}\ and\ \bibinfo {author} {\bibfnamefont {R.~W.}\ \bibnamefont
  {Spekkens}},\ }\href {\doibase 10.1103/PhysRevA.94.052324} {\bibfield
  {journal} {\bibinfo  {journal} {Phys. Rev. A}\ }\textbf {\bibinfo {volume}
  {94}},\ \bibinfo {pages} {052324} (\bibinfo {year} {2016})}\BibitemShut
  {NoStop}%
\bibitem [{\citenamefont {Yadin}\ \emph {et~al.}(2016)\citenamefont {Yadin},
  \citenamefont {Ma}, \citenamefont {Girolami}, \citenamefont {Gu},\ and\
  \citenamefont {Vedral}}]{sio}%
  \BibitemOpen
  \bibfield  {author} {\bibinfo {author} {\bibfnamefont {B.}~\bibnamefont
  {Yadin}}, \bibinfo {author} {\bibfnamefont {J.}~\bibnamefont {Ma}}, \bibinfo
  {author} {\bibfnamefont {D.}~\bibnamefont {Girolami}}, \bibinfo {author}
  {\bibfnamefont {M.}~\bibnamefont {Gu}}, \ and\ \bibinfo {author}
  {\bibfnamefont {V.}~\bibnamefont {Vedral}},\ }\href {\doibase
  10.1103/PhysRevX.6.041028} {\bibfield  {journal} {\bibinfo  {journal} {Phys.
  Rev. X}\ }\textbf {\bibinfo {volume} {6}},\ \bibinfo {pages} {041028}
  (\bibinfo {year} {2016})}\BibitemShut {NoStop}%
\bibitem [{\citenamefont {de~Vicente}\ and\ \citenamefont
  {Streltsov}(2016)}]{gc}%
  \BibitemOpen
  \bibfield  {author} {\bibinfo {author} {\bibfnamefont {J.~I.}\ \bibnamefont
  {de~Vicente}}\ and\ \bibinfo {author} {\bibfnamefont {A.}~\bibnamefont
  {Streltsov}},\ }\href {\doibase 10.1088/1751-8121/50/4/045301} {\bibfield
  {journal} {\bibinfo  {journal} {J. Phys. A: Math. Theor.}\ }\textbf {\bibinfo
  {volume} {50}},\ \bibinfo {pages} {045301} (\bibinfo {year}
  {2016})}\BibitemShut {NoStop}%
\bibitem [{\citenamefont {Winter}\ and\ \citenamefont {Yang}(2016)}]{op1}%
  \BibitemOpen
  \bibfield  {author} {\bibinfo {author} {\bibfnamefont {A.}~\bibnamefont
  {Winter}}\ and\ \bibinfo {author} {\bibfnamefont {D.}~\bibnamefont {Yang}},\
  }\href {\doibase 10.1103/PhysRevLett.116.120404} {\bibfield  {journal}
  {\bibinfo  {journal} {Phys. Rev. Lett.}\ }\textbf {\bibinfo {volume} {116}},\
  \bibinfo {pages} {120404} (\bibinfo {year} {2016})}\BibitemShut {NoStop}%
\bibitem [{\citenamefont {Chitambar}\ and\ \citenamefont
  {Gour}(2016{\natexlab{b}})}]{pio}%
  \BibitemOpen
  \bibfield  {author} {\bibinfo {author} {\bibfnamefont {E.}~\bibnamefont
  {Chitambar}}\ and\ \bibinfo {author} {\bibfnamefont {G.}~\bibnamefont
  {Gour}},\ }\href {\doibase 10.1103/PhysRevLett.117.030401} {\bibfield
  {journal} {\bibinfo  {journal} {Phys. Rev. Lett.}\ }\textbf {\bibinfo
  {volume} {117}},\ \bibinfo {pages} {030401} (\bibinfo {year}
  {2016}{\natexlab{b}})}\BibitemShut {NoStop}%
\bibitem [{\citenamefont {Yu}\ \emph {et~al.}(2016)\citenamefont {Yu},
  \citenamefont {Zhang}, \citenamefont {Xu},\ and\ \citenamefont
  {Tong}}]{alter}%
  \BibitemOpen
  \bibfield  {author} {\bibinfo {author} {\bibfnamefont {X.-D.}\ \bibnamefont
  {Yu}}, \bibinfo {author} {\bibfnamefont {D.-J.}\ \bibnamefont {Zhang}},
  \bibinfo {author} {\bibfnamefont {G.~F.}\ \bibnamefont {Xu}}, \ and\ \bibinfo
  {author} {\bibfnamefont {D.~M.}\ \bibnamefont {Tong}},\ }\href {\doibase
  10.1103/PhysRevA.94.060302} {\bibfield  {journal} {\bibinfo  {journal} {Phys.
  Rev. A}\ }\textbf {\bibinfo {volume} {94}},\ \bibinfo {pages} {060302}
  (\bibinfo {year} {2016})}\BibitemShut {NoStop}%
\bibitem [{\citenamefont {Rana}\ \emph
  {et~al.}(2016{\natexlab{a}})\citenamefont {Rana}, \citenamefont {Parashar},\
  and\ \citenamefont {Lewenstein}}]{trace}%
  \BibitemOpen
  \bibfield  {author} {\bibinfo {author} {\bibfnamefont {S.}~\bibnamefont
  {Rana}}, \bibinfo {author} {\bibfnamefont {P.}~\bibnamefont {Parashar}}, \
  and\ \bibinfo {author} {\bibfnamefont {M.}~\bibnamefont {Lewenstein}},\
  }\href {\doibase 10.1103/PhysRevA.93.012110} {\bibfield  {journal} {\bibinfo
  {journal} {Phys. Rev. A}\ }\textbf {\bibinfo {volume} {93}},\ \bibinfo
  {pages} {012110} (\bibinfo {year} {2016}{\natexlab{a}})}\BibitemShut
  {NoStop}%
\bibitem [{\citenamefont {Yao}\ \emph {et~al.}(2015)\citenamefont {Yao},
  \citenamefont {Xiao}, \citenamefont {Ge},\ and\ \citenamefont {Sun}}]{multi}%
  \BibitemOpen
  \bibfield  {author} {\bibinfo {author} {\bibfnamefont {Y.}~\bibnamefont
  {Yao}}, \bibinfo {author} {\bibfnamefont {X.}~\bibnamefont {Xiao}}, \bibinfo
  {author} {\bibfnamefont {L.}~\bibnamefont {Ge}}, \ and\ \bibinfo {author}
  {\bibfnamefont {C.~P.}\ \bibnamefont {Sun}},\ }\href {\doibase
  10.1103/PhysRevA.92.022112} {\bibfield  {journal} {\bibinfo  {journal} {Phys.
  Rev. A}\ }\textbf {\bibinfo {volume} {92}},\ \bibinfo {pages} {022112}
  (\bibinfo {year} {2015})}\BibitemShut {NoStop}%
\bibitem [{\citenamefont {Shao}\ \emph {et~al.}(2015)\citenamefont {Shao},
  \citenamefont {Xi}, \citenamefont {Fan},\ and\ \citenamefont {Li}}]{ft}%
  \BibitemOpen
  \bibfield  {author} {\bibinfo {author} {\bibfnamefont {L.-H.}\ \bibnamefont
  {Shao}}, \bibinfo {author} {\bibfnamefont {Z.}~\bibnamefont {Xi}}, \bibinfo
  {author} {\bibfnamefont {H.}~\bibnamefont {Fan}}, \ and\ \bibinfo {author}
  {\bibfnamefont {Y.}~\bibnamefont {Li}},\ }\href {\doibase
  10.1103/PhysRevA.91.042120} {\bibfield  {journal} {\bibinfo  {journal} {Phys.
  Rev. A}\ }\textbf {\bibinfo {volume} {91}},\ \bibinfo {pages} {042120}
  (\bibinfo {year} {2015})}\BibitemShut {NoStop}%
\bibitem [{\citenamefont {Zhang}\ \emph
  {et~al.}(2017{\natexlab{b}})\citenamefont {Zhang}, \citenamefont {Chen},
  \citenamefont {Li}, \citenamefont {Fei},\ and\ \citenamefont {Long}}]{cpt}%
  \BibitemOpen
  \bibfield  {author} {\bibinfo {author} {\bibfnamefont {H.-J.}\ \bibnamefont
  {Zhang}}, \bibinfo {author} {\bibfnamefont {B.}~\bibnamefont {Chen}},
  \bibinfo {author} {\bibfnamefont {M.}~\bibnamefont {Li}}, \bibinfo {author}
  {\bibfnamefont {S.-M.}\ \bibnamefont {Fei}}, \ and\ \bibinfo {author}
  {\bibfnamefont {G.-L.}\ \bibnamefont {Long}},\ }\href {\doibase
  10.1088/0253-6102/67/2/166} {\bibfield  {journal} {\bibinfo  {journal}
  {Commun. Theor. Phys.}\ }\textbf {\bibinfo {volume} {67}},\ \bibinfo {pages}
  {166} (\bibinfo {year} {2017}{\natexlab{b}})}\BibitemShut {NoStop}%
\bibitem [{\citenamefont {Rana}\ \emph
  {et~al.}(2016{\natexlab{b}})\citenamefont {Rana}, \citenamefont {Parashar},\
  and\ \citenamefont {Lewenstein}}]{lp}%
  \BibitemOpen
  \bibfield  {author} {\bibinfo {author} {\bibfnamefont {S.}~\bibnamefont
  {Rana}}, \bibinfo {author} {\bibfnamefont {P.}~\bibnamefont {Parashar}}, \
  and\ \bibinfo {author} {\bibfnamefont {M.}~\bibnamefont {Lewenstein}},\
  }\href {\doibase 10.1103/PhysRevA.93.012110} {\bibfield  {journal} {\bibinfo
  {journal} {Phys. Rev. A}\ }\textbf {\bibinfo {volume} {93}},\ \bibinfo
  {pages} {012110} (\bibinfo {year} {2016}{\natexlab{b}})}\BibitemShut
  {NoStop}%
\bibitem [{\citenamefont {Yu}(2017)}]{i1}%
  \BibitemOpen
  \bibfield  {author} {\bibinfo {author} {\bibfnamefont {C.-s.}\ \bibnamefont
  {Yu}},\ }\href {\doibase 10.1103/PhysRevA.95.042337} {\bibfield  {journal}
  {\bibinfo  {journal} {Phys. Rev. A}\ }\textbf {\bibinfo {volume} {95}},\
  \bibinfo {pages} {042337} (\bibinfo {year} {2017})}\BibitemShut {NoStop}%
\bibitem [{\citenamefont {Zhao}\ and\ \citenamefont {Yu}(2018)}]{i2}%
  \BibitemOpen
  \bibfield  {author} {\bibinfo {author} {\bibfnamefont {H.}~\bibnamefont
  {Zhao}}\ and\ \bibinfo {author} {\bibfnamefont {C.-S.}\ \bibnamefont {Yu}},\
  }\href {\doibase 10.1038/s41598-017-18692-1} {\bibfield  {journal} {\bibinfo
  {journal} {Sci. Rep.}\ }\textbf {\bibinfo {volume} {8}},\ \bibinfo {pages}
  {299} (\bibinfo {year} {2018})}\BibitemShut {NoStop}%
\bibitem [{\citenamefont {Luo}\ and\ \citenamefont
  {Sun}(2017{\natexlab{a}})}]{i3}%
  \BibitemOpen
  \bibfield  {author} {\bibinfo {author} {\bibfnamefont {S.}~\bibnamefont
  {Luo}}\ and\ \bibinfo {author} {\bibfnamefont {Y.}~\bibnamefont {Sun}},\
  }\href {\doibase 10.1103/PhysRevA.96.022130} {\bibfield  {journal} {\bibinfo
  {journal} {Phys. Rev. A}\ }\textbf {\bibinfo {volume} {96}},\ \bibinfo
  {pages} {022130} (\bibinfo {year} {2017}{\natexlab{a}})}\BibitemShut
  {NoStop}%
\bibitem [{\citenamefont {Rastegin}(2016)}]{i4}%
  \BibitemOpen
  \bibfield  {author} {\bibinfo {author} {\bibfnamefont {A.~E.}\ \bibnamefont
  {Rastegin}},\ }\href {\doibase 10.1103/PhysRevA.93.032136} {\bibfield
  {journal} {\bibinfo  {journal} {Phys. Rev. A}\ }\textbf {\bibinfo {volume}
  {93}},\ \bibinfo {pages} {032136} (\bibinfo {year} {2016})}\BibitemShut
  {NoStop}%
\bibitem [{\citenamefont {Girolami}(2014{\natexlab{b}})}]{i5}%
  \BibitemOpen
  \bibfield  {author} {\bibinfo {author} {\bibfnamefont {D.}~\bibnamefont
  {Girolami}},\ }\href {\doibase 10.1103/PhysRevLett.113.170401} {\bibfield
  {journal} {\bibinfo  {journal} {Phys. Rev. Lett.}\ }\textbf {\bibinfo
  {volume} {113}},\ \bibinfo {pages} {170401} (\bibinfo {year}
  {2014}{\natexlab{b}})}\BibitemShut {NoStop}%
\bibitem [{\citenamefont {Luo}\ and\ \citenamefont
  {Sun}(2017{\natexlab{b}})}]{i6}%
  \BibitemOpen
  \bibfield  {author} {\bibinfo {author} {\bibfnamefont {S.}~\bibnamefont
  {Luo}}\ and\ \bibinfo {author} {\bibfnamefont {Y.}~\bibnamefont {Sun}},\
  }\href {\doibase 10.1103/PhysRevA.96.022136} {\bibfield  {journal} {\bibinfo
  {journal} {Phys. Rev. A}\ }\textbf {\bibinfo {volume} {96}},\ \bibinfo
  {pages} {022136} (\bibinfo {year} {2017}{\natexlab{b}})}\BibitemShut
  {NoStop}%
\bibitem [{\citenamefont {Napoli}\ \emph {et~al.}(2016)\citenamefont {Napoli},
  \citenamefont {Bromley}, \citenamefont {Cianciaruso}, \citenamefont {Piani},
  \citenamefont {Johnston},\ and\ \citenamefont {Adesso}}]{op3}%
  \BibitemOpen
  \bibfield  {author} {\bibinfo {author} {\bibfnamefont {C.}~\bibnamefont
  {Napoli}}, \bibinfo {author} {\bibfnamefont {T.~R.}\ \bibnamefont {Bromley}},
  \bibinfo {author} {\bibfnamefont {M.}~\bibnamefont {Cianciaruso}}, \bibinfo
  {author} {\bibfnamefont {M.}~\bibnamefont {Piani}}, \bibinfo {author}
  {\bibfnamefont {N.}~\bibnamefont {Johnston}}, \ and\ \bibinfo {author}
  {\bibfnamefont {G.}~\bibnamefont {Adesso}},\ }\href {\doibase
  10.1103/PhysRevLett.116.150502} {\bibfield  {journal} {\bibinfo  {journal}
  {Phys. Rev. Lett.}\ }\textbf {\bibinfo {volume} {116}},\ \bibinfo {pages}
  {150502} (\bibinfo {year} {2016})}\BibitemShut {NoStop}%
\bibitem [{\citenamefont {Uhlmann}(1998)}]{eroof}%
  \BibitemOpen
  \bibfield  {author} {\bibinfo {author} {\bibfnamefont {A.}~\bibnamefont
  {Uhlmann}},\ }\href {\doibase 10.1023/A:1009664331611} {\bibfield  {journal}
  {\bibinfo  {journal} {Open Systems {\&} Information Dynamics}\ }\textbf
  {\bibinfo {volume} {5}},\ \bibinfo {pages} {209} (\bibinfo {year}
  {1998})}\BibitemShut {NoStop}%
\bibitem [{\citenamefont {Bennett}\ \emph {et~al.}(1996)\citenamefont
  {Bennett}, \citenamefont {DiVincenzo}, \citenamefont {Smolin},\ and\
  \citenamefont {Wootters}}]{eroof1}%
  \BibitemOpen
  \bibfield  {author} {\bibinfo {author} {\bibfnamefont {C.~H.}\ \bibnamefont
  {Bennett}}, \bibinfo {author} {\bibfnamefont {D.~P.}\ \bibnamefont
  {DiVincenzo}}, \bibinfo {author} {\bibfnamefont {J.~A.}\ \bibnamefont
  {Smolin}}, \ and\ \bibinfo {author} {\bibfnamefont {W.~K.}\ \bibnamefont
  {Wootters}},\ }\href {\doibase 10.1103/PhysRevA.54.3824} {\bibfield
  {journal} {\bibinfo  {journal} {Phys. Rev. A}\ }\textbf {\bibinfo {volume}
  {54}},\ \bibinfo {pages} {3824} (\bibinfo {year} {1996})}\BibitemShut
  {NoStop}%
\bibitem [{\citenamefont {Vidal}(2000)}]{monotone}%
  \BibitemOpen
  \bibfield  {author} {\bibinfo {author} {\bibfnamefont {G.}~\bibnamefont
  {Vidal}},\ }\href {\doibase 10.1080/09500340008244048} {\bibfield  {journal}
  {\bibinfo  {journal} {J. Mod. Opt.}\ }\textbf {\bibinfo {volume} {47}},\
  \bibinfo {pages} {355} (\bibinfo {year} {2000})}\BibitemShut {NoStop}%
\bibitem [{\citenamefont {Yu}\ and\ \citenamefont {Song}(2009)}]{roof4}%
  \BibitemOpen
  \bibfield  {author} {\bibinfo {author} {\bibfnamefont {C.-s.}\ \bibnamefont
  {Yu}}\ and\ \bibinfo {author} {\bibfnamefont {H.-s.}\ \bibnamefont {Song}},\
  }\href {\doibase 10.1103/PhysRevA.80.022324} {\bibfield  {journal} {\bibinfo
  {journal} {Phys. Rev. A}\ }\textbf {\bibinfo {volume} {80}},\ \bibinfo
  {pages} {022324} (\bibinfo {year} {2009})}\BibitemShut {NoStop}%
\bibitem [{\citenamefont {Yuan}\ \emph {et~al.}(2015)\citenamefont {Yuan},
  \citenamefont {Zhou}, \citenamefont {Cao},\ and\ \citenamefont {Ma}}]{roof2}%
  \BibitemOpen
  \bibfield  {author} {\bibinfo {author} {\bibfnamefont {X.}~\bibnamefont
  {Yuan}}, \bibinfo {author} {\bibfnamefont {H.}~\bibnamefont {Zhou}}, \bibinfo
  {author} {\bibfnamefont {Z.}~\bibnamefont {Cao}}, \ and\ \bibinfo {author}
  {\bibfnamefont {X.}~\bibnamefont {Ma}},\ }\href {\doibase
  10.1103/PhysRevA.92.022124} {\bibfield  {journal} {\bibinfo  {journal} {Phys.
  Rev. A}\ }\textbf {\bibinfo {volume} {92}},\ \bibinfo {pages} {022124}
  (\bibinfo {year} {2015})}\BibitemShut {NoStop}%
\bibitem [{\citenamefont {Qi}\ \emph {et~al.}(2017)\citenamefont {Qi},
  \citenamefont {Gao},\ and\ \citenamefont {Yan}}]{roof3}%
  \BibitemOpen
  \bibfield  {author} {\bibinfo {author} {\bibfnamefont {X.}~\bibnamefont
  {Qi}}, \bibinfo {author} {\bibfnamefont {T.}~\bibnamefont {Gao}}, \ and\
  \bibinfo {author} {\bibfnamefont {F.}~\bibnamefont {Yan}},\ }\href {\doibase
  10.1088/1751-8121/aa7638} {\bibfield  {journal} {\bibinfo  {journal} {J.
  Phys. A: Math. Theor.}\ }\textbf {\bibinfo {volume} {50}},\ \bibinfo {pages}
  {285301} (\bibinfo {year} {2017})}\BibitemShut {NoStop}%
\bibitem [{\citenamefont {Du}\ \emph {et~al.}(2015)\citenamefont {Du},
  \citenamefont {Bai},\ and\ \citenamefont {Qi}}]{oto1}%
  \BibitemOpen
  \bibfield  {author} {\bibinfo {author} {\bibfnamefont {S.}~\bibnamefont
  {Du}}, \bibinfo {author} {\bibfnamefont {Z.}~\bibnamefont {Bai}}, \ and\
  \bibinfo {author} {\bibfnamefont {X.}~\bibnamefont {Qi}},\ }\href@noop {}
  {\bibfield  {journal} {\bibinfo  {journal} {Quantum Information and
  Computation}\ }\textbf {\bibinfo {volume} {15}} (\bibinfo {year}
  {2015})}\BibitemShut {NoStop}%
\bibitem [{\citenamefont {Liu}\ \emph {et~al.}(2017)\citenamefont {Liu},
  \citenamefont {Hu},\ and\ \citenamefont {Lloyd}}]{qrt3}%
  \BibitemOpen
  \bibfield  {author} {\bibinfo {author} {\bibfnamefont {Z.-W.}\ \bibnamefont
  {Liu}}, \bibinfo {author} {\bibfnamefont {X.}~\bibnamefont {Hu}}, \ and\
  \bibinfo {author} {\bibfnamefont {S.}~\bibnamefont {Lloyd}},\ }\href
  {\doibase 10.1103/PhysRevLett.118.060502} {\bibfield  {journal} {\bibinfo
  {journal} {Phys. Rev. Lett.}\ }\textbf {\bibinfo {volume} {118}},\ \bibinfo
  {pages} {060502} (\bibinfo {year} {2017})}\BibitemShut {NoStop}%
\bibitem [{\citenamefont {Bu}\ \emph {et~al.}(2017)\citenamefont {Bu},
  \citenamefont {Singh}, \citenamefont {Fei}, \citenamefont {Pati},\ and\
  \citenamefont {Wu}}]{mre}%
  \BibitemOpen
  \bibfield  {author} {\bibinfo {author} {\bibfnamefont {K.}~\bibnamefont
  {Bu}}, \bibinfo {author} {\bibfnamefont {U.}~\bibnamefont {Singh}}, \bibinfo
  {author} {\bibfnamefont {S.-M.}\ \bibnamefont {Fei}}, \bibinfo {author}
  {\bibfnamefont {A.~K.}\ \bibnamefont {Pati}}, \ and\ \bibinfo {author}
  {\bibfnamefont {J.}~\bibnamefont {Wu}},\ }\href {\doibase
  10.1103/PhysRevLett.119.150405} {\bibfield  {journal} {\bibinfo  {journal}
  {Phys. Rev. Lett.}\ }\textbf {\bibinfo {volume} {119}},\ \bibinfo {pages}
  {150405} (\bibinfo {year} {2017})}\BibitemShut {NoStop}%
\bibitem [{\citenamefont {Liu}\ \emph {et~al.}(2019)\citenamefont {Liu},
  \citenamefont {Bu},\ and\ \citenamefont {Takagi}}]{oneshot}%
  \BibitemOpen
  \bibfield  {author} {\bibinfo {author} {\bibfnamefont {Z.-W.}\ \bibnamefont
  {Liu}}, \bibinfo {author} {\bibfnamefont {K.}~\bibnamefont {Bu}}, \ and\
  \bibinfo {author} {\bibfnamefont {R.}~\bibnamefont {Takagi}},\ }\href
  {\doibase 10.1103/PhysRevLett.123.020401} {\bibfield  {journal} {\bibinfo
  {journal} {Phys. Rev. Lett.}\ }\textbf {\bibinfo {volume} {123}},\ \bibinfo
  {pages} {020401} (\bibinfo {year} {2019})}\BibitemShut {NoStop}%
\bibitem [{\citenamefont {Nielsen}\ and\ \citenamefont {Chuang}(2002)}]{qi2}%
  \BibitemOpen
  \bibfield  {author} {\bibinfo {author} {\bibfnamefont {M.~A.}\ \bibnamefont
  {Nielsen}}\ and\ \bibinfo {author} {\bibfnamefont {I.}~\bibnamefont
  {Chuang}},\ }\href {\doibase 10.1119/1.1463744} {\bibfield  {journal}
  {\bibinfo  {journal} {Am. J. Phys.}\ }\textbf {\bibinfo {volume} {70}},\
  \bibinfo {pages} {558} (\bibinfo {year} {2002})}\BibitemShut {NoStop}%
\bibitem [{\citenamefont {Horodecki}\ and\ \citenamefont
  {Oppenheim}(2013)}]{qrt2}%
  \BibitemOpen
  \bibfield  {author} {\bibinfo {author} {\bibfnamefont {M.}~\bibnamefont
  {Horodecki}}\ and\ \bibinfo {author} {\bibfnamefont {J.}~\bibnamefont
  {Oppenheim}},\ }\href {\doibase 10.1142/S0217979213450197} {\bibfield
  {journal} {\bibinfo  {journal} {Int. J. Mod. Phys. B}\ }\textbf {\bibinfo
  {volume} {27}},\ \bibinfo {pages} {1345019} (\bibinfo {year}
  {2013})}\BibitemShut {NoStop}%
\bibitem [{\citenamefont {Peng}\ \emph {et~al.}(2016)\citenamefont {Peng},
  \citenamefont {Jiang},\ and\ \citenamefont {Fan}}]{decom}%
  \BibitemOpen
  \bibfield  {author} {\bibinfo {author} {\bibfnamefont {Y.}~\bibnamefont
  {Peng}}, \bibinfo {author} {\bibfnamefont {Y.}~\bibnamefont {Jiang}}, \ and\
  \bibinfo {author} {\bibfnamefont {H.}~\bibnamefont {Fan}},\ }\href {\doibase
  10.1103/PhysRevA.93.032326} {\bibfield  {journal} {\bibinfo  {journal} {Phys.
  Rev. A}\ }\textbf {\bibinfo {volume} {93}},\ \bibinfo {pages} {032326}
  (\bibinfo {year} {2016})}\BibitemShut {NoStop}%
\bibitem [{\citenamefont {Plenio}\ and\ \citenamefont
  {Virmani}(2014)}]{Plenio2014}%
  \BibitemOpen
  \bibfield  {author} {\bibinfo {author} {\bibfnamefont {M.~B.}\ \bibnamefont
  {Plenio}}\ and\ \bibinfo {author} {\bibfnamefont {S.~S.}\ \bibnamefont
  {Virmani}},\ }\enquote {\bibinfo {title} {An introduction to entanglement
  theory},}\ in\ \href {\doibase 10.1007/978-3-319-04063-9_8} {\emph {\bibinfo
  {booktitle} {Quantum Information and Coherence}}},\ \bibinfo {editor} {edited
  by\ \bibinfo {editor} {\bibfnamefont {E.}~\bibnamefont {Andersson}}\ and\
  \bibinfo {editor} {\bibfnamefont {P.}~\bibnamefont {{\"O}hberg}}}\ (\bibinfo
  {publisher} {Springer International Publishing},\ \bibinfo {address} {Cham},\
  \bibinfo {year} {2014})\ pp.\ \bibinfo {pages} {173--209}\BibitemShut
  {NoStop}%
\bibitem [{\citenamefont {Rana}\ \emph {et~al.}(2017)\citenamefont {Rana},
  \citenamefont {Parashar}, \citenamefont {Winter},\ and\ \citenamefont
  {Lewenstein}}]{ne}%
  \BibitemOpen
  \bibfield  {author} {\bibinfo {author} {\bibfnamefont {S.}~\bibnamefont
  {Rana}}, \bibinfo {author} {\bibfnamefont {P.}~\bibnamefont {Parashar}},
  \bibinfo {author} {\bibfnamefont {A.}~\bibnamefont {Winter}}, \ and\ \bibinfo
  {author} {\bibfnamefont {M.}~\bibnamefont {Lewenstein}},\ }\href {\doibase
  10.1103/PhysRevA.96.052336} {\bibfield  {journal} {\bibinfo  {journal} {Phys.
  Rev. A}\ }\textbf {\bibinfo {volume} {96}},\ \bibinfo {pages} {052336}
  (\bibinfo {year} {2017})}\BibitemShut {NoStop}%
\bibitem [{\citenamefont {Plenio}(2005)}]{ln}%
  \BibitemOpen
  \bibfield  {author} {\bibinfo {author} {\bibfnamefont {M.~B.}\ \bibnamefont
  {Plenio}},\ }\href {\doibase 10.1103/PhysRevLett.95.090503} {\bibfield
  {journal} {\bibinfo  {journal} {Phys. Rev. Lett.}\ }\textbf {\bibinfo
  {volume} {95}},\ \bibinfo {pages} {090503} (\bibinfo {year}
  {2005})}\BibitemShut {NoStop}%
\bibitem [{\citenamefont {Zhu}\ \emph {et~al.}(2017)\citenamefont {Zhu},
  \citenamefont {Ma}, \citenamefont {Cao}, \citenamefont {Fei},\ and\
  \citenamefont {Vedral}}]{oto2}%
  \BibitemOpen
  \bibfield  {author} {\bibinfo {author} {\bibfnamefont {H.}~\bibnamefont
  {Zhu}}, \bibinfo {author} {\bibfnamefont {Z.}~\bibnamefont {Ma}}, \bibinfo
  {author} {\bibfnamefont {Z.}~\bibnamefont {Cao}}, \bibinfo {author}
  {\bibfnamefont {S.-M.}\ \bibnamefont {Fei}}, \ and\ \bibinfo {author}
  {\bibfnamefont {V.}~\bibnamefont {Vedral}},\ }\href {\doibase
  10.1103/PhysRevA.96.032316} {\bibfield  {journal} {\bibinfo  {journal} {Phys.
  Rev. A}\ }\textbf {\bibinfo {volume} {96}},\ \bibinfo {pages} {032316}
  (\bibinfo {year} {2017})}\BibitemShut {NoStop}%
\bibitem [{\citenamefont {Du}\ \emph {et~al.}(2019)\citenamefont {Du},
  \citenamefont {Bai},\ and\ \citenamefont {Qi}}]{tran2}%
  \BibitemOpen
  \bibfield  {author} {\bibinfo {author} {\bibfnamefont {S.}~\bibnamefont
  {Du}}, \bibinfo {author} {\bibfnamefont {Z.}~\bibnamefont {Bai}}, \ and\
  \bibinfo {author} {\bibfnamefont {X.}~\bibnamefont {Qi}},\ }\href {\doibase
  10.1103/PhysRevA.100.032313} {\bibfield  {journal} {\bibinfo  {journal}
  {Phys. Rev. A}\ }\textbf {\bibinfo {volume} {100}},\ \bibinfo {pages}
  {032313} (\bibinfo {year} {2019})}\BibitemShut {NoStop}%
\end{thebibliography}%

\end{document}